\documentclass[preprint,aps,nofootinbib,preprintnumbers,amsmath,amssymb,11pt]{revtex4}
\usepackage{epsfig}
\usepackage{float}
\graphicspath{ {images/} }
\usepackage{ytableau}
\usepackage{mwe}
\usepackage{amsmath}
\usepackage{multirow}
\usepackage{slashed}
\usepackage{xcolor}
\usepackage{amssymb}
\usepackage{color}
\definecolor{darkblue}{rgb}{0.,0.,0.4}
\definecolor{darkred}{rgb}{0.5,0.,0.}
\definecolor{BlueViolet}{RGB}{138,43,226}
\definecolor{SkyBlue}{RGB}{30,144,255}
\definecolor{DarkGreen}{RGB}{0,100,0}
\usepackage[colorlinks=true,linkcolor=blue,citecolor=blue]{hyperref}
\usepackage{amsfonts}
\usepackage{graphicx}
\usepackage{dcolumn}
\usepackage{bm}
\usepackage{natbib}

\newcommand{\be}{\begin{equation}}
\newcommand{\ee}{\end{equation}}
\newcommand{\bea}{\begin{eqnarray}}
\newcommand{\eea}{\end{eqnarray}}

\def\im{{{\rm i}}}
\newcommand{\junchen}[1]{ { \color{red} \footnotesize (\textsf{JR}) \textsf{\textsl{#1}} }}

\begin{document}


\title{Scalar conformal field theories from lattice systems}

\author{\vspace*{1 cm}Junchen Rong}
\email{junchenrong@gmail.com}
\affiliation{\vspace*{1 cm}Institut des Hautes \'Etudes Scientifiques, 91440 Bures-sur-Yvette, France
\\ }
\affiliation{
\\ CPHT, CNRS, \'Ecole Polytechnique, Institut Polytechnique de Paris, 91120 Palaiseau, France
\vspace*{2 cm}
}

\begin{abstract}
\vspace*{1 cm}
We discuss scalar conformal field theories (CFTs) that can be realized in structural phase transitions.
The Landau condition and Lifshitz condition are reviewed, which are necessary conditions for a structural phase transition to be second order. 
We also review the perturbative analysis in $4-\epsilon$ expansion of the corresponding Landau actions, which were already analyzed thoroughly in the 80s. 
By identifying the global symmetries of these fixed points, it turns out that in perturbation theory only 6 different CFTs can be realized by commensurate structural phase transitions. 
\textbf{Updated in version 2:} We discuss how to classify all the phases of a Landau theory using the computer algebra system GAP.
We also discuss the fully packed quantum loop model on the triangular lattice, on which the Cubic CFT is realized.

This is a lecture note based on a series of talks given by the author. 
The note aims to bridge the gap between condensed matter physicists and conformal field theorists. 
The note will be further updated in the future.

\end{abstract}

\vspace*{3 cm}

\maketitle

\newpage


\tableofcontents
\section{Structure phase transition}
\subsection{Introduction}
Structural phase transitions describe the change of crystal structure as the critical temperature is crossed \cite{cowley1980structural}.
It is an interesting phenomenon that connects material science, statistical physics, quantum field theory, and group theory. 
Given a crystal, there exist transformations (translation, rotation, reflection with respect to a certain plane, etc) that leave the crystal structure invariant. These transformations form a group structure, which is called the crystallographic ``space group''.
Three-dimensional Euclidean space $R^3$ is invariant under the so-called Euclidean group $E(3)$. 
The group $E(3)$ contains the following group elements 
\begin{align}
\text{Translation:} & \quad x^i \rightarrow x^i+a^i,  &\quad i\in {1,2,3},\nonumber\\
\text{rotation+reflection:} & \quad x^i \rightarrow \sum_{j=1}^{3} R_i{}^j x^j, &\quad  R_i{}^j \in O(3),
\end{align}
and their combinations. 
The matrix $R_i^{j}$ is an $3\times 3$ orthogonal matrix, that is, $R_i{}^j$ is an element of the three dimensional orthogonal group $O(3)$. 
Notice that the Euclidean group leaves invariant the Cartesian distance 
$\sum_{i=1}^{3} (x^i-y^i)^2$.
A space group is a subgroup of $E(3)$ that leaves the discrete crystal structure invariant. 
The classification of space groups (and the study of their representations, group-subgroup relation and etc) is one of the most successful programs in mathematical physics \cite{bradley2010mathematical,aroyo2013international}. 
The structural phase transitions can be understood through the theory of spontaneous symmetry breaking. 
At higher temperatures, the crystal has a structure that preserves a space group $G$. 
At a temperature below the phase transition temperature, the new crystal structure preserves a different space group $H$ which is a subgroup of $G$~\footnote{For some materials, the high-temperature phase may have smaller symmetry than the low-temperature phase.  Such exotic inverted transitions were observed in Rochelle's salt \cite{weinberg1974gauge,jona1962ferroelectric}.}.
There exist two types of phase transitions, continuous and discontinuous phase transitions. 
The terms ``continuous" and ``discontinuous" refer to whether the order parameter changes continuously (see Section \ref{Landautheory}) when the phase transition happens. 
One ``surprising'' result of a continuous phase transition is that it is possible for completely different physical systems to have the same universal critical behavior near the critical temperature $T_c$. 
For example, the liquid-gas phase transition at the critical point and the lattice Ising model at its critical temperature have the exact same critical behavior (see Section \ref{RGfluctuation}). 
Because of this, continuous phase transitions can be classified into so-called ``universality classes'', according to their critical behavior.

According to the modern theory of continuous phase transitions, the critical behavior near the critical temperature $T_c$ is completely fixed by the behavior of thermal fluctuations exactly at $T_c$. 
The latter can then be described by a special type of quantum field theory called the conformal field theory (CFT). 
As compared to regular quantum field theories, CFTs preserve extra symmetry than the Euclidean group $E(3)$. 
This gives field theorists extra mileage in studying them. 
In particular, the critical behavior of many two-dimensional phase transitions can be solved exactly using CFT techniques \cite{Belavin:1984vu}. 
Recently, the development of the conformal bootstrap technique \cite{Poland:2018epd} has greatly improved our understanding of many conformal field theories in $d\geq2$, especially the CFTs that can be realized by structural phase transitions (see Table \ref{universality}). 
Conformal bootstrap, together with Monte Carlo simulation (see for example \cite{landau2021guide}), are currently among the most successful methods in studying critical phenomena.  
Inspired by these new developments in CFT research, we review here the early literature on structural phase transitions. 
There already exist nice textbooks and reviews on structural phase transitions, such as \cite{toledano1987landau,cowley1980structural}. 
We will however give a brief review of the subject, emphasising the relation to conformal field theories so as to bridge the gap between the two fields of research.
In particular, for a structural phase transition to be second order, it satisfies the following necessary conditions
\begin{itemize}
    \item the group-subgroup relation,
    \item the Landau condition,
    \item the (weak) Lifshitz condition,
    \item and stability under renormalization group flow.
\end{itemize}
We will discuss these conditions in the following sections. 
We need to warn the readers that these conditions are based on either mean field theory or perturbation theory arguments. 
When the non-perturbative effect are taken into account, phase transitions violating these conditions may also be second order. 
We will mention some counterexamples in future sections. 
After examining the early literature on structural phase transitions (which were conveniently reported in Table 12 of the book \cite{stokes1988isotropy}) and identify the global symmetry groups of the perturbative fixed points, it turns out that in perturbation theory, only six different CFTs can be realized by structural phase transitions, which are summarized in Table \ref{universality}. 
\begin{table}[h]
\begin{tabular}{|l|l|l|}
\hline
No. & Name        & Images                                      \\ \hline\hline
1   & Ising       & A2a                                         \\ \hline
2   & XY          & B4a,  B6b,  B8a,  B12a,  B12b,  B24a        \\ \hline
3   & $N$=3 Cubic & C24a, C24c, C48a                            \\ \hline
4   & XY$^2$      & D32e,  D64a, D64b, D64d, D72b, D128a, D144a \\ \hline
5   & $N$=4 Cubic & D192a, D192c, D384a                         \\ \hline
6   & XY$^3$      & E96k, E192j, E768b, E768c, E1536a           \\ \hline
\end{tabular}
\caption{All perturbative critical universality classes which can be realized in structural phase transitions. See Section \ref{Crystaluniversalities} for details. 
We make this table by identifying the symmetry groups of the perturbative fixed points reported in Table 12 of the book \cite{stokes1988isotropy}. 
The table is based on perturbation theory in $4-\epsilon$ dimensions. The non-perturbative effect may change the result.}\label{universality}
\end{table}
A fully non-perturbative result is still unavailable (see Section \ref{Crystaluniversalities}).

\subsection{Landau theory for spontaneous symmetry breaking}\label{Landautheory}
\subsubsection{Invariant polynomials and the Landau condition}
We will now explain Landau's theory of spontaneous symmetry breaking.
Let us start with the Ising model.
The Ising model is given by the following Hamiltonian 
\be{}\label{IsingHam}
H_{\text{Ising}}=-J \sum_{\langle ij\rangle} \sigma_i \sigma_j.
\ee
Here $\langle ij \rangle$ denotes the nearest neighbor sites on the lattice. 
On each site, $\sigma_{i}=\pm 1$.
Clearly, the Hamiltonian of the Ising model preserves the $Z_2$ symmetry, under which all the spins flip sign 
\be{}
\sigma_i\rightarrow -\sigma_i.
\ee
The partition function of the Ising model is given by 
\be{}
Z(T)=\sum_{\sigma_i} e^{-\beta H}, \quad \text{with}\quad  \beta=\frac{1}{k_B T}.
\ee
Here $k_B$ is the Boltzmann constant, which we set to be equal to 1 for simplicity. The expectation value of the spin operator is given by 
\be{}
\langle \sigma_x \rangle_T =\frac{\sum_{\sigma_i} \sigma_x e^{-\beta H}}{Z(T)}.
\ee
\begin{figure}[ht]
\includegraphics[width=8cm]{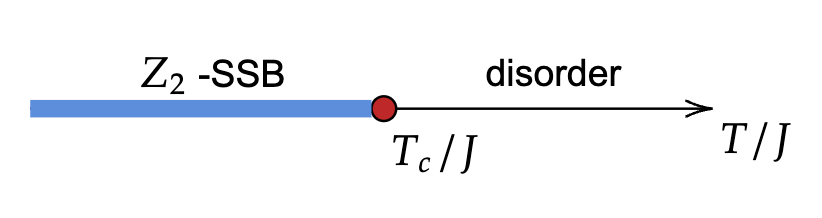}
\caption{The phase diagram of the Ising model. $T>T_c$ is the disordered phase, while the $T<T_c$ is the ordered phase with spontaneously broken $Z_2$ symmetry.}\label{IsingPhaseDiagram}
\end{figure}
Onsager's famous solution of the Ising model on two dimensional square lattice \cite{PhysRev.65.117} tells us that the phase diagram is given by Fig. \ref{IsingPhaseDiagram}.
When $T>T_c$, the Ising model is in the disordered phase, in which
\be{}
\langle \sigma_x \rangle_T=0.
\ee
When $T>T_c$, the Ising model is in the so called ordered phase, in which the $Z_2$ symmetry is spontaneously broken,
\be{}
\langle \sigma_x \rangle_T=\pm v.
\ee
Here $v$ is a constant that depends on the temperature. 
The system has two degenerate vacua, $\langle \sigma_x \rangle_T=\pm v$ that are related to each other by the $Z_2$ transformation.
Even though the Hamiltonian of the Ising model is symmetric under $Z_2$, the vacuum is not.
This phenomenon is called spontaneous symmetry breaking.

\begin{figure}[ht]
\includegraphics[width=8cm]{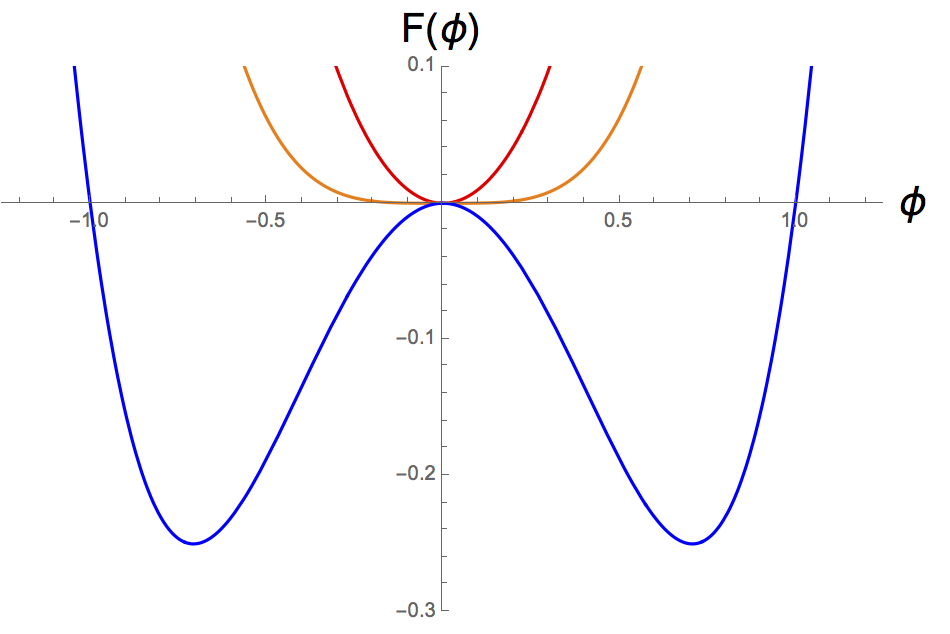}
 \caption{The free energy $F(\phi)=a \phi^2+\phi^4$. The red, orange, and blue curves correspond to $a$=1, 0, and -1 respectively.}
\label{phi4}
\end{figure}

The Landau theory is a generic theory about spontaneous symmetry breaking. In general, the theory depends on two factors, the symmetry group $G$ and the irreducible representation that the order parameter transforms in. (Here we assume a single order parameter.) 
Let us first consider the simplest case, in which the symmetry group is $Z_2$, which is also the symmetry group of the Ising model.
We can easily write down the free energy function that is invariant under the $Z_2$ operation $\phi\rightarrow -\phi$,
\be{}
F(\phi)=a \phi^2+\lambda \phi^4+ \lambda_6 \phi^6+\cdots.
\ee
The coupling constants $a$, $\lambda$, and $\lambda_6$, in general, depend on the temperature. In the case of the Ising model, $\phi$ should be understood as the vacuum expectation value $\langle \sigma_x \rangle_T$. The location of the minima of this function depends on the sign of $a$ (assuming $\lambda>0$ and $\lambda_6\geq 0$), see Figure \ref{phi4}.
The phase transition happens precisely when $a$ changes sign. 
The parameter $a$ depends on the temperature.
Near the critical temperature, one can perform a linear expansion to get $a\propto (T-T_c)$. 
When $T>T_c$, the minimum of the free energy potential is located at $$\phi=0.$$ 
When $T<T_c$, on the other hand, there are two minima located at 
$$\phi=\pm v,$$
with the constant $v$ depending on the temperature. 
The phase diagram of the Landau theory therefore agrees with the Ising model.
When $T$ is slightly below the critical temperature $T_c$, the spontaneous magnetization can be approximated by 
\be{}
\phi\propto (T_c-T)^{\beta}. 
\ee
The specific heat has the following power law behavior 
\be{}
C_P=-\frac{\partial^2 F}{\partial T^2}\propto |T-T_c|^{-\alpha}.
\ee
The constants $\alpha$ and $\beta$ are called critical exponents, which characterize the universality class that the second-order phase transition belongs to. 
The Landau theory gives us the mean-field theory value $\alpha=0$ and $\beta=1/2$. 
In general, $\alpha$ and $\beta$ will be different from their mean field theory values.

Universality means that there exist different physical systems, whose critical exponents at the second-order phase transition are the same. 
For example, the critical point of the liquid-gas phase transition and the Curie point of the magnetization phase transition belong to the same universality class.

Landau argued that for a phase transition to be second order, the symmetry of the low-temperature phase $H$ must be a subgroup of the high-temperature group $G$. 
This also means that for two phases preserving $G$ and $G'$ respectively, if $G$ is not a subgroup of $G'$ and vice versa, the two phases can only be connected by a discontinuous phase transition. 
The ``group-subgroup relation'' between phases is, therefore, a necessary condition for second-order phase transition~\footnote{Quantum phase transitions beyond the Landau(-Ginzburg-Wilson) paradigm has been proposed~\cite{senthil2004deconfined}. 
In these cases, phases with two incompatible symmetries can be connected by second-order phase transitions.}. 

\begin{figure}[ht]
\includegraphics[width=8cm]{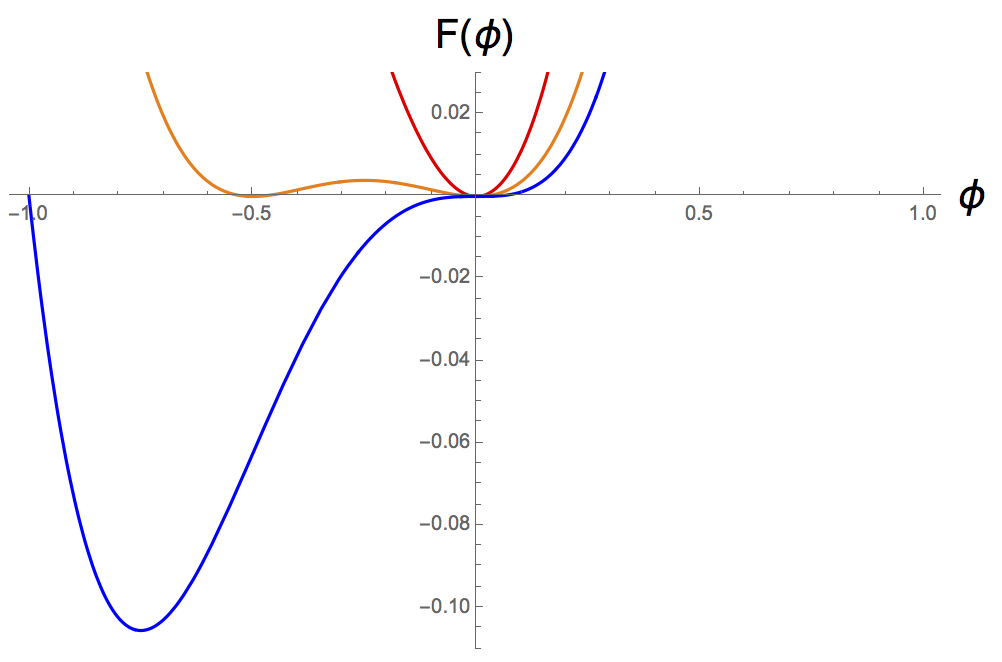}
\caption{The free energy $F(\phi)=a \phi^2+\phi^3+\phi^4$. The red, orange, and blue curves correspond to $a_2$=1, 1/4, and 0 respectively. 
At $a$=0, the barrier between the meta-stable (at $\phi=0$) and the true vacuum at $\phi<0$ disappears, the first-order phase transition happens.}
\label{phi3}
\end{figure}

Landau also argued that if a phase transition is second order, the free energy function can not contain cubic ($\phi^3$) terms. 
This is sometimes called the Landau condition.
An illustration is given in Fig. \ref{phi3}.
In most cases, the symmetry groups we will encounter in structural phase transitions are finite groups. 
To construct the Landau theory which preserves a finite group $G$, it will be nice to know how many polynomial terms can appear in each degree.  
The Molien function does precisely this job.  
For a finite group G, the Molien function of its representation $\rho(g)$ is
\be
M(z)=\frac{1}{|G|}\sum_{g\in G}\frac{1}{Det[\mathbf{1}-z \rho(g)]}.
\ee
It is a generating function that counts the number of invariant polynomials of a certain degree. 
Here $|G|$ is the order of the group G, $\rho(g)$ is a matrix, which is also a representation of G.
In the Taylor expansion of the  Molien function $M(z)$ around $z=0$, the coefficient of $z^n$ indicates the number of invariant polynomials of degree $n$.
See, for example, \cite{stokes1988isotropy,Liendo:2021wpo}, where the Molien function was used to study effective actions. 
As explained in \cite{stokes1988isotropy}, the Molien function can be written as 
\be\label{molienfunctiongenaric}
M(z)=\frac{\beta_0+\beta_1 z+\ldots \beta_m z^m}{(1-z)^{\alpha_1}(1-z^2)^{\alpha_2}\ldots (1-z^n)^{\alpha_n}}.
\ee
A generic invariant polynomial of the group G can be written as
\be\label{generalpoly}
P=\sum_{j=1}^{m} P_j^{(r)} K_j(P_1^{(b)},P_2^{(b)},\ldots P_n^{(b)})
\ee
The  $P_i^{(b)}$ polynomials with $i=1\ldots n$ are called the ``basic'' invariant polynomials.
Basic invariant polynomials contribute to the denominator of the Molien function \eqref{molienfunctiongenaric}.
The numbers $\alpha_n$ count the number of basic invariant polynomials of degree $n$.
The $P_j^{(r)}$ polynomials with $j=1\ldots m$ are called the ``relative'' invariant polynomials. 
Relative invariant polynomials contribute to the numerator of the Molien function \eqref{molienfunctiongenaric}.
For a finite group, the number of relative and basic invariant polynomials is finite. 
The function $K_j$ is itself a polynomial, with $P_i^{(b)}$'s as its variables. 
Since basis polynomials $P_i^{(b)}$'s are invariant polynomials, $K_j$ is also invariant. 
This explains the denominator of \eqref{molienfunctiongenaric}. 
Compared to basic invariant polynomials, relative polynomials have a special property that  $(P_j^{(r)})^{n}$ with $n\geq 2$ is not independent, it can be re-expressed as linear combinations of terms in \eqref{generalpoly}. This explains the numerator of \eqref{molienfunctiongenaric}. 
Let us consider as an example the symmetric group $S_4$. 
The Molien function of the standard representation of $S_4$ (the permutation group of four elements) is 
\be
M(z)=\frac{1}{(1-z^2)(1-z^3)(1-z^4)}=1+z^2+z^3+\ldots.
\ee
This data can be conveniently obtained using the following GAP~\cite{GAP4} code:

\smallskip
\qquad\qquad\qquad\qquad\qquad\qquad \parbox{0.4\textwidth}{grp:=SymmetricGroup(4);\\ tbl:=CharacterTable(grp);\\ psi:=Irr(tbl);\\ MolienSeries(psi[4]);}
\smallskip\\\\
The first line of the code specifies which group we wish to consider. 
The second line calculates the character table of the group.
The third generates a list ``psi'' which contains all the irreps of the group.
The fourth line calculated the Molien function of the 4th irrep, which is the standard representation.
The GAP system has a library called ``Smallgroup'', which allows us to easily deal with finite groups with orders less than 2000. 
One can easily obtain the character table, a matrix representation of the generators of the finite group and even the Molien functions using this library.
To get an explicit form of the invariant polynomials, we need to first get a matrix representation of the generators. The group $S_4$ is generated by the permutation
\be
(1,2,3,4) \quad \text{and} \quad (1,2).
\ee
Here we are using the cycle notation for elements of the symmetric group. 
By typing ``IrreducibleRepresentationsDixon(grp, psi[4]: unitary);'' in GAP, we get 
\be
(1,2,3,4)=\left(
\begin{array}{ccc}
 0 & -\frac{1}{\sqrt{3}} & -\sqrt{\frac{2}{3}} \\
 \frac{1}{\sqrt{3}} & -\frac{2}{3} & \frac{\sqrt{2}}{3} \\
 \sqrt{\frac{2}{3}} & \frac{\sqrt{2}}{3} & -\frac{1}{3} \\
\end{array}
\right), \quad \text{and}\quad (1,2)=\left(
\begin{array}{ccc}
 \frac{1}{2} & \frac{1}{2 \sqrt{3}} & -\sqrt{\frac{2}{3}} \\
 \frac{1}{2 \sqrt{3}} & \frac{5}{6} & \frac{\sqrt{2}}{3} \\
 -\sqrt{\frac{2}{3}} & \frac{\sqrt{2}}{3} & -\frac{1}{3} \\
\end{array}
\right).
\ee
To calculate the explicit form of the invariant polynomials, we can use, for example, the built-in functions of Mathematica. 
Take the degree three invariant polynomials as an example, we first use ``m1=KroneckerProduct[g1,g1,g1]'' to construct a $3^{degree}\times 3^{degree}$ matrix from the generators.
Here g1 is one of the generators, it is a $3\times 3$ matrix. 
The matrix ``m1'' tells us how the generators act on the tensor product space $V\otimes V\otimes V$, suppose $V$ is the standard irrep of $S_4$. 
The command ``NullSpace[m1-IdentityMatrix[27]]'' then calculates the 27-dimensional vectors corresponding to the invariant tensors. 
One should then solve for linear combinations of these null vectors which are also invariant under the action of ``KroneckerProduct[g2,g2,g2]'' (where g2 is the second generator).
Convert the vectors back to the tensorial basis we get the invariant tensors, which are equivalent to the invariant polynomials.

The three basic invariant polynomials of the standard representation of the symmetric group $S_4$ with degree-two, three, and four are given by 
\bea
I_2 (x_i)&=& \sum_i x_i^2,\\
I_3(x_i)&=&-\frac{x_2^3}{\sqrt{2}}-\frac{3}{2} x_3 x_2^2+x_3^3+\frac{3}{2} x_1^2 \left(\sqrt{2} x_2-x_3\right),\\
I_{4}(x_i)&=&d_{ijm}d_{klm}x_ix_jx_kx_l.
\eea
(Summation over repeated indices are understood.) Here the invariant tensor $d_{ijk}$ is defined as 
\be\label{tensor}
d_{ijk}=\frac{\partial^3 I_3}{ \partial x_i\partial x_j\partial x_k}.
\ee
A lattice model that preserves the $S_4$ symmetry is the 4-state Potts model, which is a generalization of the Ising model, allowing the spins to take 4 values instead. 
The Hamiltonian is 
\be{}
H_{\text{Potts}}=-J\sum_{\langle ij \rangle} \delta_{s_i,s_j}.
\ee
Here $\delta_{s_i,s_j}$ is the Kronecker delta function, which equals 1 when $s_i=s_j$ and 0 otherwise. 
The spins $s_i$ can take four values 0,1,2, and 3. 
The numerical simulation of this model shows a phase diagram that is very similar to the phase diagram of the Ising model. 
The difference is that the low-temperature phase is the symmetry-breaking phase of $S_4$.
Now we can write down the effective action of the 4-state Potts model according to \eqref{generalpoly}. 
The leading terms are 
\be
F(\phi)= a_2 I_2(\phi_i)+a_3 I_3(\phi_i)+a_{4,a} I_{4}(\phi_i)+ a_{4,b} \left(I_{2}(\phi_i)\right)^2+\ldots
\ee
(For an explicit form of the effective action of N-state Potts models with generic N, see 
\cite{Zia:1975ha}.) 
Clearly, the effective action of the four-state Potts model contains a cubic term, this is consistent with the $z^3$ term in the Molien series. 
According to Landau's argument, this transition should be first order. 
This is indeed true in three dimensions (see for example \cite{bazavov2008phase}). 
In two dimensions, however, the 4-state Potts model goes through a second-order phase transition, because 
the effect of thermal fluctuation is stronger in two dimensions \cite{baxter1973potts,di1987relations,duminil2017continuity}. 
In particular, the cubic operator $I_3(\phi_i)$ gets strongly renormalized and becomes irrelevant, see Section \ref{RGfluctuation}. 
In Section~\ref{phasesofLandau}, we will discuss how to classify the phases of a generic Landau action.

The Molien function of the standard representation of $A_4$, the alternating group on four elements, is 
\be
M(z)=\frac{1+z^6}{(1-z^2)(1-z^3)(1-z^4)}.
\ee
Notice the Molien function of $S_4$ and $A_4$ are exactly the same up to $z^5$, the difference starts at $z^6$ order. 
The group $A_4$ is a subgroup of $S_4$, which consists only of the even permutations of four elements.
The standard irrep of $S_4$, when branching into $A_4$, remains irreducible. 
The group $A_4$ shall preserve more invariant polynomials than $S_4$. 
Suppose that we want to explicitly break the global symmetry group of the Landau theory from $S_4$ to $A_4$. 
From the discussion above, this means we will have to introduce $\phi^6$ terms. 
In two dimensions, this operator is irrelevant at the 4-state Potts model fixed point. This means that a UV model with $A_4$ symmetry will also be in the 4-state Potts model universality class. In general, the symmetry $G$ of the second-order phase transition point, which is described by conformal field theories (see Section \ref{RGfluctuation}), can be bigger than the symmetry $H$ of the lattice model.
As long as all operators that are singlet of $H$ but carrying non-trivial quantum numbers of $G$ are irrelevant.

\subsubsection{Irreps of space groups and images}\label{landautheoryandimage}
Bravais lattices are 3D lattices defined as a set of vectors 
\be{}
\Vec{R}=\sum_{i=1}^{3} n_i \Vec{a}_i, \quad \text{with}\quad n_i\in \mathbb{Z}.
\ee
The three linearly independent vectors $\Vec{a}_i$ are called the primitive lattice vectors, and they define the so-called unit cell. 
The volume of the unit cell is given by
\be
\Omega=\epsilon_{\mu\nu\rho}a_1^{\mu} a_2^{\nu} a_3^{\rho}.
\ee
We can define primitive reciprocal lattice vectors as 
\be
{b}_1^{\mu}=\frac{2\pi}{\Omega}\epsilon_{\mu\nu\rho}a_2^{\nu} a_3^{\rho},\quad {b}_2^{\mu}=\frac{2\pi}{\Omega} \epsilon_{\mu\nu\rho}a_3^{\nu} a_1^{\rho},\quad {b}_3^{\mu}=\frac{2\pi}{\Omega} \epsilon_{\mu\nu\rho}a_1^{\nu} a_2^{\rho}.
\ee
The reciprocal lattice consist of vectors given by $\Vec{K}=\sum_{i} n_i \Vec{b}_i$, with $n_i$ again integers.
Primitive lattice vectors and reciprocal lattice vectors satisfy 
\be\label{abrelations}
\Vec{a}_i\cdot \Vec{b}_j=2\pi \delta_{ij}, \quad \text{and}\quad \sum_{i} a_i^{\mu} b_i^{\nu}=2\pi \delta^{\mu\nu}.
\ee
The group that leaves the lattice invariant is called the space group $G$ of the lattice. 
$G$ is a subgroup of the three-dimensional Euclidean group $E(3)$. 
Clearly, translation by a primitive lattice vector leaves the lattice invariant. 
The set of all such translations forms the transnational group $\mathcal{T}$, which is an Abelian normal subgroup of the space group.
The group $G$ also contains the subgroup of rotations and reflections that leave the lattice invariant, which we will denote as $\mathcal{P}$. 
$\mathcal{P}$ is also called a crystallographic point group, which is a subgroup of the three-dimensional orthogonal group $O(3)$~\footnote{For a generic space group containing glide mirrors and screw axes, the definition of point group is more subtle. 
A generic space group element is $\{g,\vec{t}\}$, which acts on a vector as 
\be{}
\{g,\vec{t}\} \vec{r}=g \cdot\vec{r}+\vec{t}\nonumber.
\ee
The point group $\mathcal{P}$ is the group of all $g$'s.
}.

For a Bravais lattice, all the lattice sites are made of the same type of atoms. 
It is also simply a tiling of the three dimensional flat space with empty unit cells. 
We can fill these cells with atoms that are different from the atoms living on the Bravais lattice sites.
These new lattices will have a space group symmetry which is the subgroup of the space group of the underlying Bravais lattice.

There are in total 32 three-dimensional crystallographic point groups, and in total 230 three-dimensional crystallographic space groups. 
The ``International Tables for Crystallography'' \cite{aroyo2013international} collects the properties of these groups, with many introductory chapters explaining the classification.
Interested readers may refer to these chapters for further information. 
The book \cite{bradley2010mathematical} is also a nice reference to study this subject. 
As we mentioned, the Landau theory of spontaneous symmetry breaking says that for a second order phase transition to happen, the symmetry $H$ of the low temperature phase must be a subgroup of the symmetry $G$ of the high temperature phase. 
To study structural phase transitions, it is therefore desirable to classify the ``group-subgroup relations'' between crystallographic space groups.
This has been done with the help of a computer, and a book containing the results was published \cite{stokes1988isotropy}.

The irreps of the translation group $\mathcal{T}$ are labeled by a momenta point $\Vec{k}$ in the unit reciprocal lattice cell, which is alternatively called the Brillouin zone. 
Bloch's theorem tells us that the eigenfunctions of the translational group can be written as \be
\rho_{\Vec{k}}(\Vec{r})=e^{i\Vec{k}\cdot\Vec{r}} u(\Vec{r}),
\ee
with $u(\vec{r})$ being a periodic function satisfying 
\be{}
u(\Vec{r}+ \Vec{a}_i)=u(\Vec{r}), \quad \text{for}\quad i=1,2,3.
\ee
Notice the periodic function can also be decomposed into Fourier modes of momenta belonging to the reciprocal lattice 
$
u(r)=\sum_{\Vec{K}} f_K e^{i\Vec{K}\cdot \Vec{r}}
$.
Under translations, the Bloch function picks a phase
\be{
}
\rho_{\Vec{k}}(\Vec{r}+\Vec{a}_i)= e^{\im \Vec{k}\cdot \Vec{a}_i}\rho_{\Vec{k}}(\Vec{r}),\quad \text{for}\quad i=1,2,3.
\ee
Notice that the Bloch function with momentum $\Vec{k}$ and $\Vec{k}+\Vec{b}_i$ are in the same irrep of the transnational group, due to \eqref{abrelations}.

\begin{figure}[h]
\includegraphics[width=12cm]{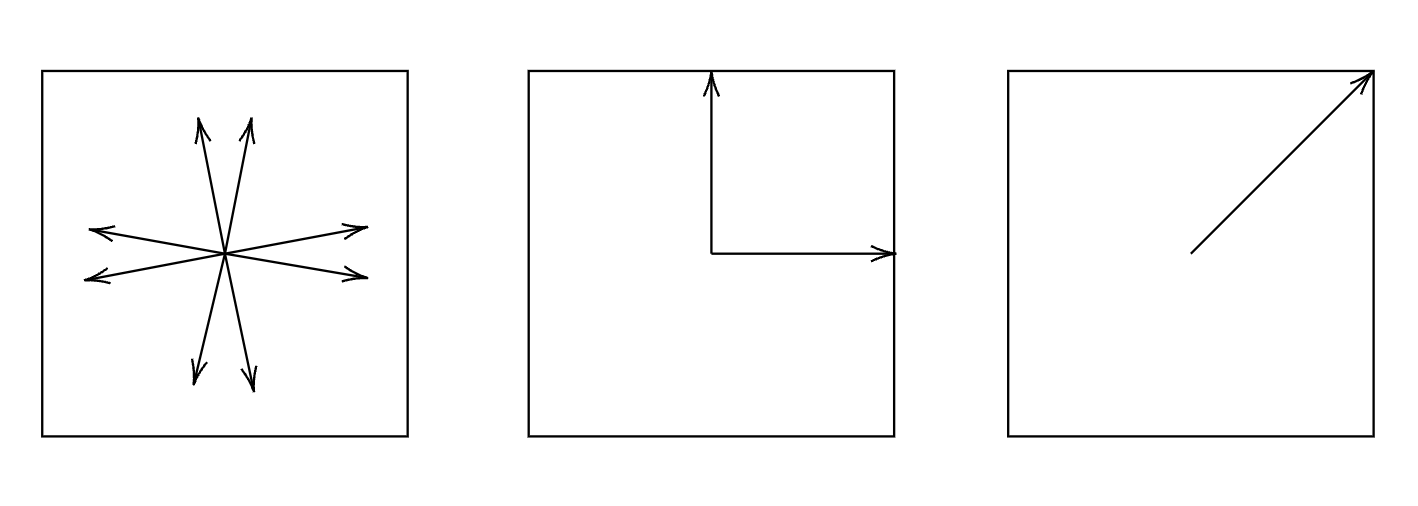}
\caption{Example stars of $\vec{k}$.}
\label{starsBZ}
\end{figure}

Starting with a point $\vec{k}$ in the Brillouin zone, the action of the point group $\cal P$ brings $\vec{k}$ to other points in the  Brillouin zone. 
The set of all these points is called the star of $\vec{k}$, denoted as $\vec{k}_*$. 
For a two-dimensional square lattice, in general, $\vec{k}_*$ contains 8 vectors. 
When $\vec{k}$ is located at some special points, such as the edge of the Brillouin zone, the $\vec{k}_*$ contains fewer vectors. 
This is because the Brillouin zone is a torus so that the vectors $\Vec{k}$ and $\Vec{k}+\Vec{b}_i$ are identical. See Fig. \ref{starsBZ}.

\begin{figure}[h]
a) \includegraphics[width=12.3cm]{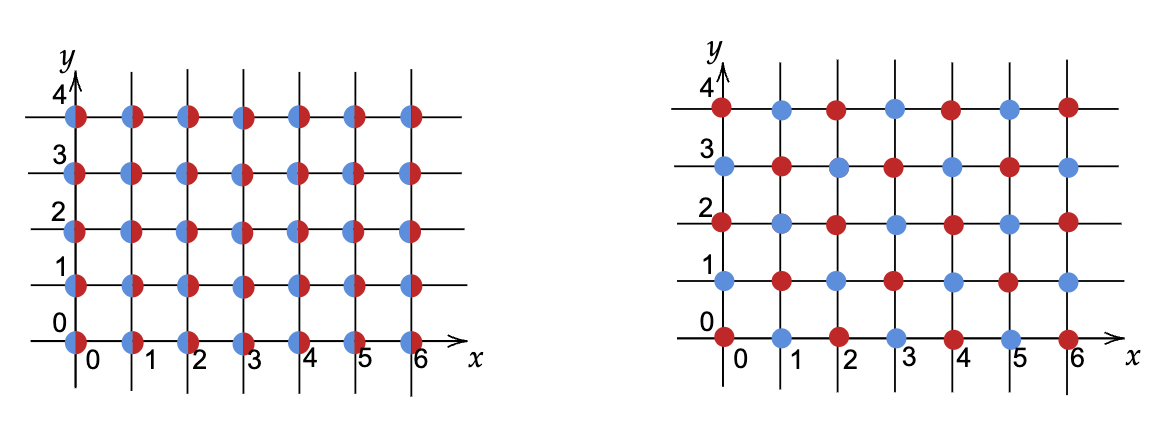}\\
b)
\includegraphics[width=4cm]{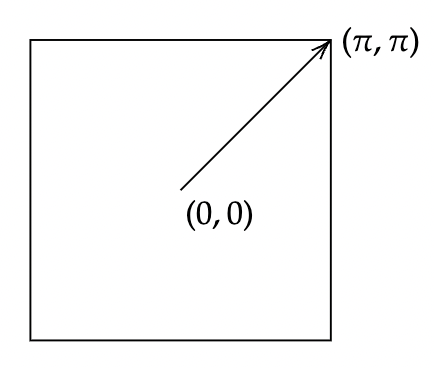}
\caption{a) The disordered and ordered phase phase of adsorbed monolayer atoms on a square lattice. The half-red and half-blue circle means that there is a 50\% probability of finding the red atom.
b) The Brillouin zone.  
The momentum of the order parameter is at the corner of the Brillouin zone. }
\label{criticalline}
\end{figure}

\begin{figure}[ht]
a)
\includegraphics[width=9cm]{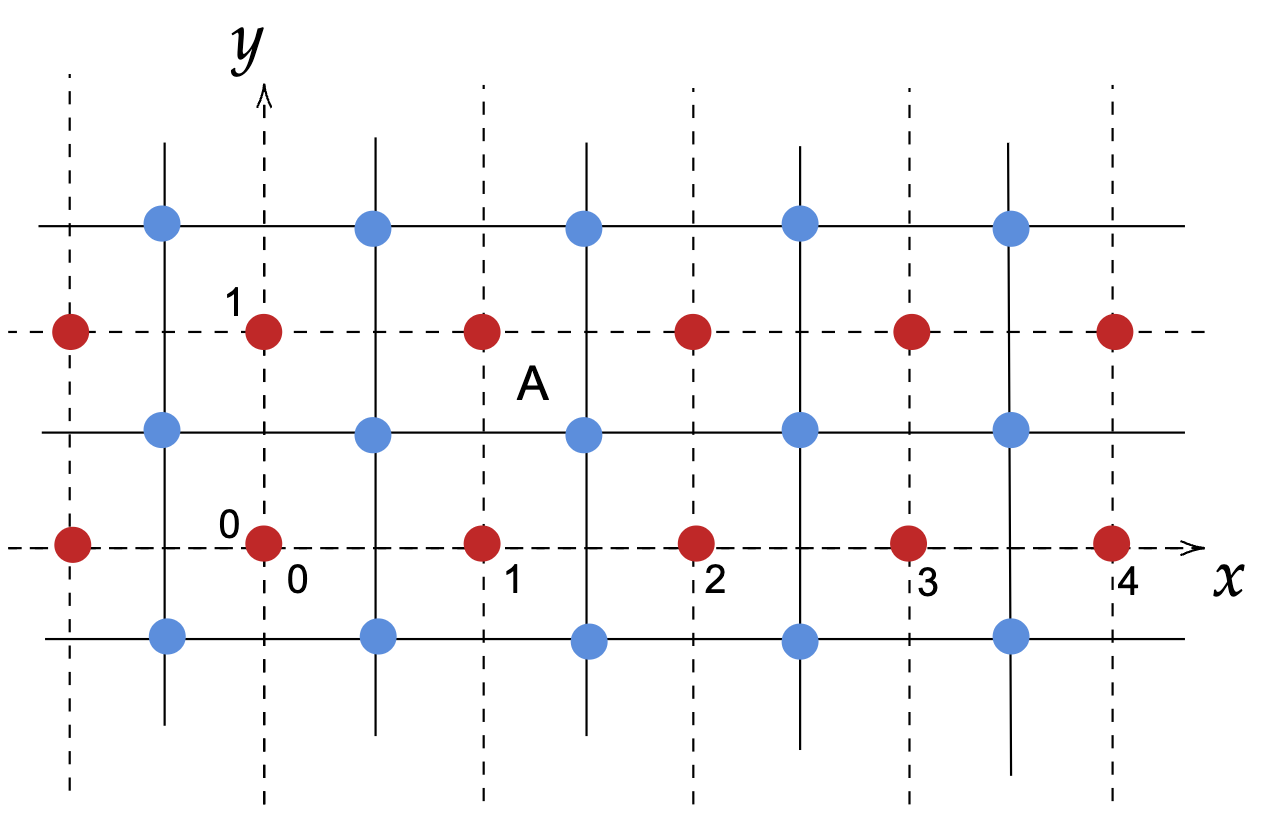}
\\
b)
\includegraphics[width=9.5cm]{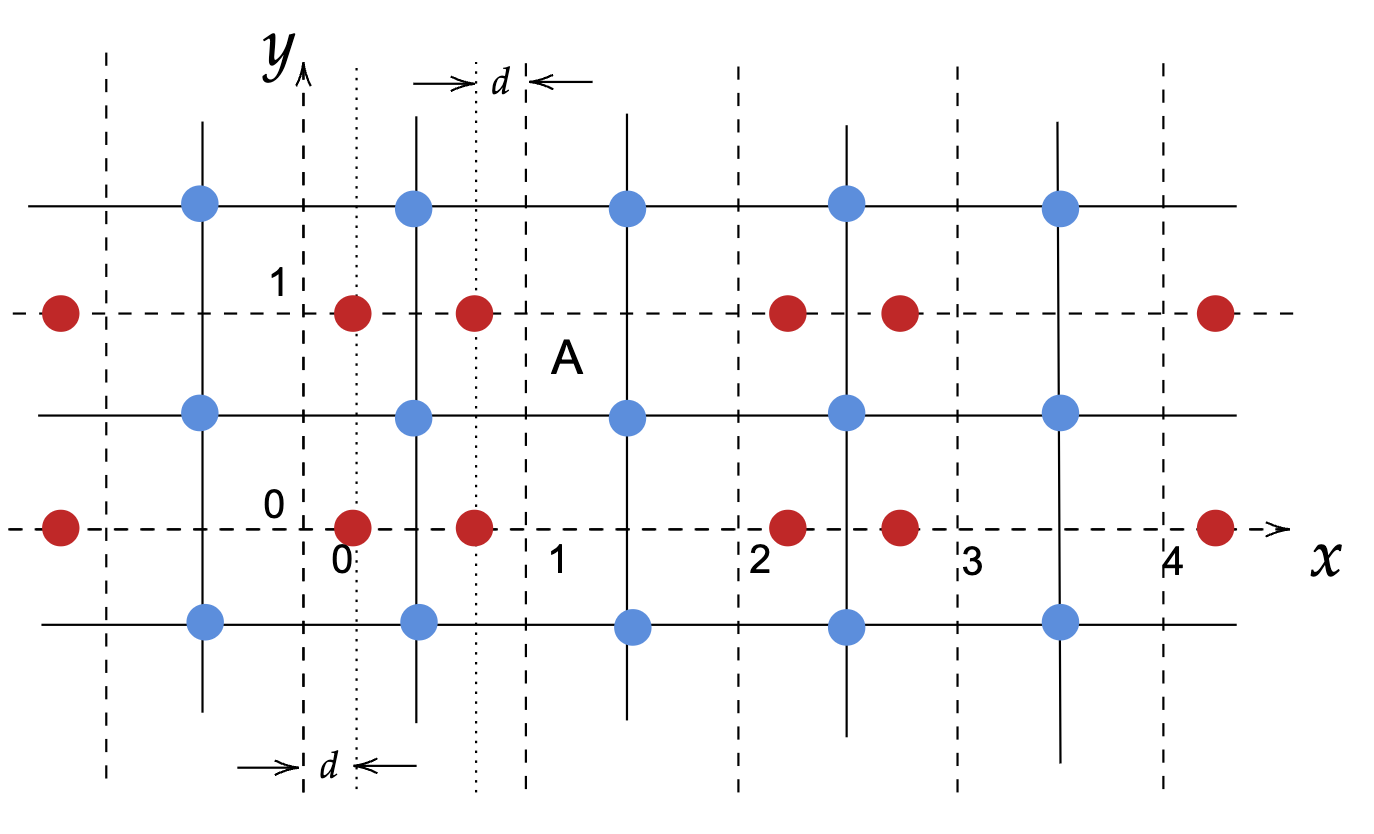}
\\
c)
\includegraphics[width=10cm]{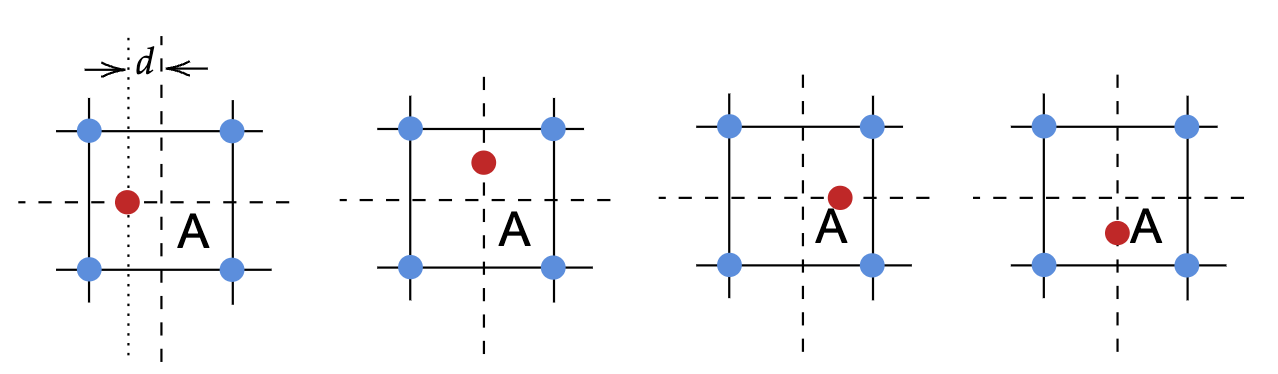}
\caption{A hypothetical displacive structural phase transition. a) In The high-temperature phase, the red and blue atoms are randomly distributed. b) The low-temperature phase. c) The four degenerate symmetry-breaking phases. The atom at block A can move in four directions when symmetry breaking happens.}
\label{displacement2d}
\end{figure}

There are two types of structural phase transitions, the order-disorder transitions, and the displacive transitions \cite{toledano1987landau}. 
In an order-disorder structural phase transition, the crystal consists of different types of atoms. 
In the high-temperature phase, different atoms can occupy the sites of the lattice sites with equal probability. In other words, the atoms are randomly distributed on the lattice.
In the low temperature phase, on the other hand, the atoms occupying the lattices form certain structures, see Fig.~\ref{criticalline}. 
In a displacive structural phase transition, the location of certain atoms changes from a more symmetrical position to a position that breaks the space group symmetry, see Fig. \ref{displacement2d}. 

For convenience, we will use two-dimensional structural phase transitions to illustrate their difference.
A type of phase transition in two dimensions that are analogous to the three dimensional structural phase transitions is the order-disorder phase transition of mono-layer atoms of molecules absorbed on the surface of certain substrate material \cite{MADEY1975304,somorjai1973low,ENGELHARDT1976591,doering1982adsorption,de1998co,stampfl1996structure}. 
The absorbed mono-layer atoms or molecules can have different phases as temperature changes. 
The phase transitions are described by the spontaneous breaking of the two-dimensional space groups, also called the wallpaper groups. 
There are only 17 of them. 
As a simple example, let us consider the order-disorder transitions of adsorbed monolayers \cite{domany1978classification,domany1979classification} on a square lattice. 
This is essentially the two-dimensional version of the order-disorder structural phase transition in three dimensions. 
At high temperatures, the absorbed atoms distribute randomly on the lattice. 
Below the critical temperature, the absorbed atoms form commensurate super-lattice structures as in Figure \ref{criticalline}.
For simplicity, we now assume the lattice constant $a=1$. That is, the primitive lattice vectors are 
\be{}
\vec{a}_1=(1,0),\quad \vec{a}_2=(0,1).
\ee
The primitive reciprocal lattice vectors are then 
\be
\vec{b}_1=(2\pi,0),\quad \vec{b}_2=(0,2\pi).
\ee
The average density of the red atoms is 
\be\label{isingdensity}
\rho(\Vec{r})=C u(\Vec{r})+\phi e^{\im \vec{k}\cdot \Vec{r}} u(\Vec{r}),\quad {\rm with}\quad\vec{k}=(\pi,\pi).
\ee
The function
\be\label{latticeDelta}
u(\Vec{r})=\sum_{\Vec{R}} \delta(\Vec{r}-\Vec{R}).
\ee
is periodic and invariant under the space group transformations. 
The high-temperature phase corresponds to 
\be{}
C=\frac{1}{2}, \quad \phi=0.
\ee
The low-temperature phase, on the other hand, corresponds to 
\be{}
C=\frac{1}{2}, \quad \phi=\frac{1}{2}.
\ee
In general, the coefficients of the two terms depend on the temperature.
The first term is a space group singlet so it does not play any role in the symmetry breaking.
We treat the second-term  as the order parameter of the phase transition.
The momentum of the order parameter lives on a special point of symmetry of the Brillouin zone, see Fig \ref{criticalline} c). 
At these points of symmetry, not all the space group elements are represented faithfully, which means that some group elements act trivially. 
The space group is generated by 90-degree rotations, reflection, and translation along the horizontal direction. 
The 90-degree rotations and reflections bring the Brillouin zone point to a new point which is equivalent to the original and, therefore acts trivially.
The translation, on the other hand, flips the sign of the order parameter as 
\be
\eta(\Vec{r}+\Vec{a}_1)=-\eta(\Vec{r}),
\ee
so that 
\be{}
\phi\rightarrow-\phi.
\ee
We have used \eqref{abrelations}. The subgroup of the space group which is faithfully represented (denoted as $G_i$, which stands for ``image group'') and the corresponding irrep of $G_i$ that the order parameter transforms in together are called the image of the space group. 
In our case,
\be
G_i=Z_2,
\ee
and the order parameter is in the odd representation of $Z_2$. The Landau effective potential is  
\be
F=a \phi^2+ \lambda \phi^4 +\cdots.
\ee
Depending on the sign of $\lambda$, the phase transition can be either first order or second order, see Fig. \ref{phi4} and Fig \ref{phi4negative}. If the transition is second order, it will be in the two-dimensional Ising universality class. 

\begin{figure}[ht]
a)\includegraphics[width=6cm]{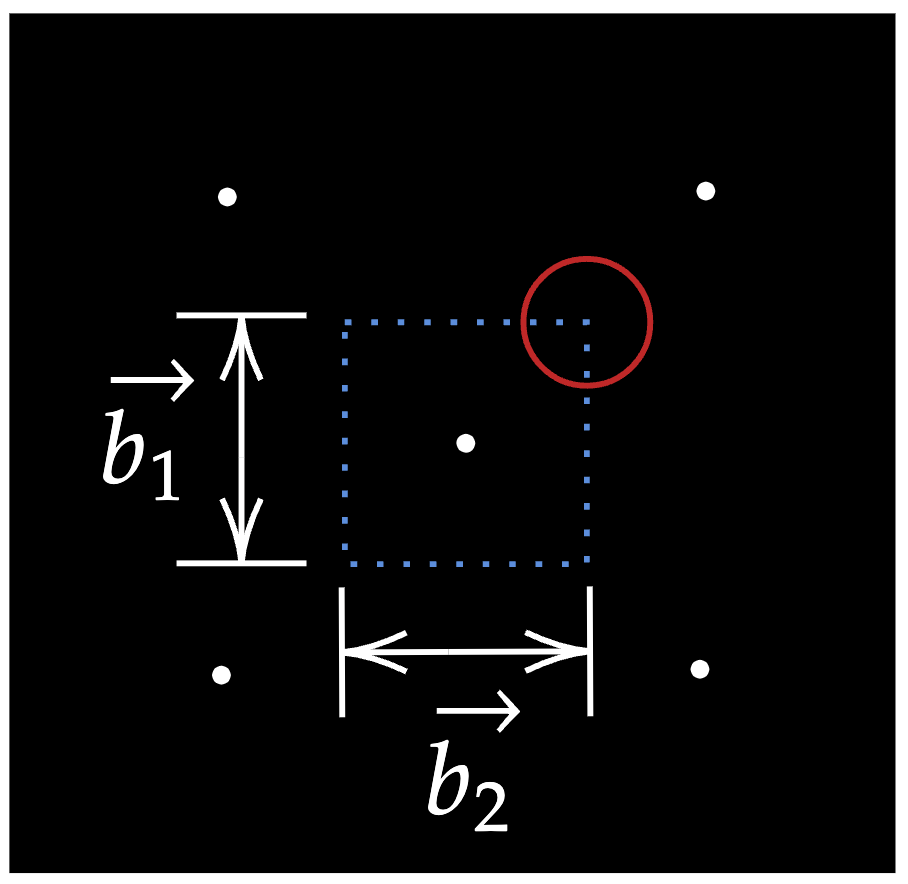}
b)
\includegraphics[width=5.65cm]{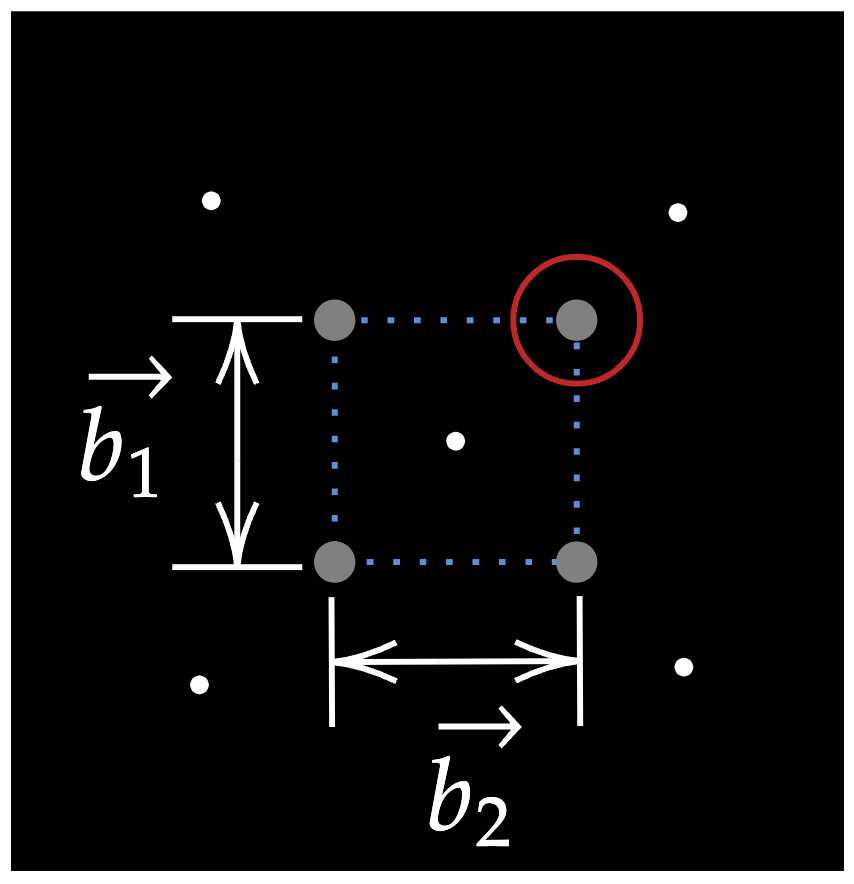}
\caption{Idealized pattern of Bragg peaks observed in LEED experiment above and below the critical temperature. The blue square corresponds to the first Brillouin zone. Measuring the integrated intensities of the diffraction beam inside the red circle gives us the critical exponents of the transition. (The bright white dots are the Bragg peak of modes that are at zero momentum.)}
\label{LEED}
\end{figure}

The order parameter can be measured in low-energy-electron-diffraction (LEED) experiments~\footnote{See \cite{cowley1980structural} for a review on experimental measurements related to 3D structural phase transitions.}.
As an example, the work of \cite{madey1975adsorption,piercy1987experimental,pfnur1990oxygen} measured the intensity of the diffraction beam at the momentum corresponding to the structure of the absorbed molecules.
From the temperature dependence of the integrated intensity, one can measure the critical exponents $\beta$ of the second order phase transition, for example, \cite{piercy1987experimental}. 
An idealized version of the pattern in the LEED experiment is given in Figure \ref{LEED}. 
Blow the critical temperature, a new Bragg peak appears at the momentum of the order parameter.

Recently, a two-dimensional displacive structural phase transition was predicted and subsequently realized in monolayer transition-metal dichalcogenides (TMDs) \cite{duerloo2014structural,wang2017structural}.
For a review of the experimental realization of two-dimensional structural phase transitions, see \cite{li2021phase}.
To understand the difference between displacive and order-disorder structural phase transitions, we consider the hypothetical structural phase transition given in Figure \ref{displacement2d}.
We again assume the lattice constant to be $a=1$. The red atoms, at low temperatures, change positions.
The displacement of the red atoms depends on the lattice sites, which can be written as 
\be\label{structuremode2}
\vec{d}(\vec{r})=\phi_1 \left(
\begin{array}{c}
 1 \\
 0 \\
\end{array}
\right) e^{i \pi x}+\phi_2 \left(
\begin{array}{c}
 0 \\
 1 \\
\end{array}
\right) e^{i \pi y}.
\ee
The phase in Fig \ref{displacement2d} b) corresponds to 
\be
\phi_1=d, \quad \phi_2=0.
\ee
By analyzing how the space group acts on the order parameter term, we can figure out how coefficients $(\phi_1,\phi_2)$ transform:
\bea{}
\text{Translation }T_x:  && \quad \phi_1 \rightarrow -\phi_1, \quad \phi_2 \rightarrow \phi_2,\nonumber\\
\text{Translation }T_y:  && \quad \phi_1 \rightarrow \phi_1, \quad \phi_2 \rightarrow -\phi_2,\nonumber\\
\text{Four-fold rotation } R_4: && \quad \phi_1 \rightarrow \phi_2,\quad \phi_2 \rightarrow -\phi_1.
\eea
The above transformations form the dihedral group $D_4$.
Notice certain elements of the space group act trivially on these order parameters, such as $(T_x)^2$. 
In other words, the space group is not faithfully represented by these order parameters. 
From the above transformations, we get
the Landau effective potential
\be\label{cubicpotential}
F=a (\phi_1^2+\phi_2^2)+ a_{4,1} (\phi_1^4+\phi_2^4)+a_{4,2} (\phi_1^2+\phi_2^2)^2 +\cdots.
\ee
The four configurations in Fig.~\ref{displacement2d} b) correspond to the four degenerate vacua of the effective potential,
\be{}
 \phi_1  = \pm d, \phi_2 =0, \quad \text{and }\quad  \phi_1 =0, \phi_2 =\pm d.
\ee
The order of the phase transition depends on the choice of $a_{4,1}$ and $a_{4,2}$.
In fact, there exists a famous lattice model with $D_4$ symmetry, which is called the Ashkin-Teller model \cite{PhysRev.64.178}. 
The phase diagram of the Ashkin-Teller model can be found in, for example, \cite{PhysRevB.22.2542}. 
The phase diagram has phases separated by a second-order phase transition line. 
Interestingly, the critical exponents of the models vary along the line. 
This is because, in two dimensions, there is a family of $c=1$ conformal field theories with $D_4$ symmetry. 
A pedagogical introduction of conformal field theories with c=1 and in particular the Ashkin-Teller models can be found in Section 8.4 of  \cite{Ginspar1988ui}.
We will not discuss the details here.

\subsubsection{Fluctuations and renormalization}\label{RGfluctuation}
\begin{figure}[ht]
a)\includegraphics[width=4cm]{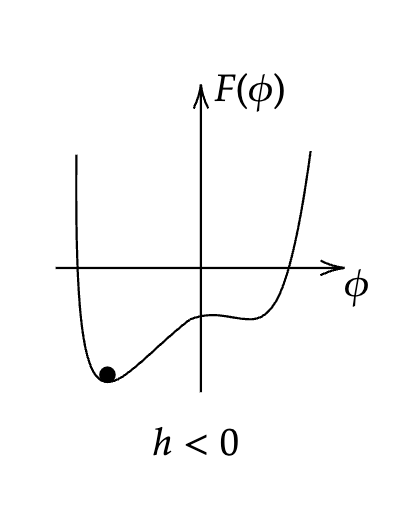}
b)\includegraphics[width=4cm]{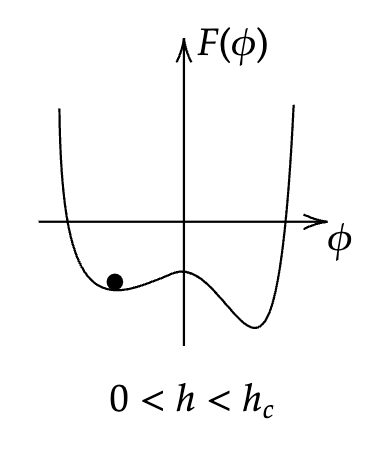}
c)\includegraphics[width=4cm]{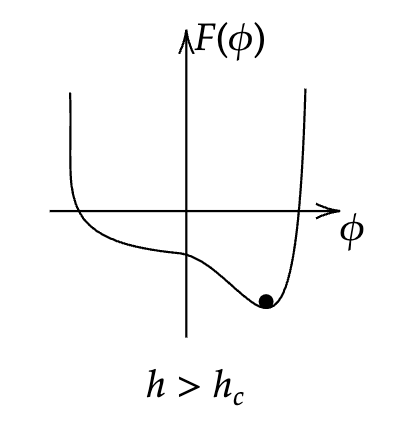}
\caption{Without fluctuations, the state can be stuck in meta-stable vacuums.}\label{Hpotential}
\end{figure}

\begin{figure}[ht]
a)\includegraphics[width=6cm]{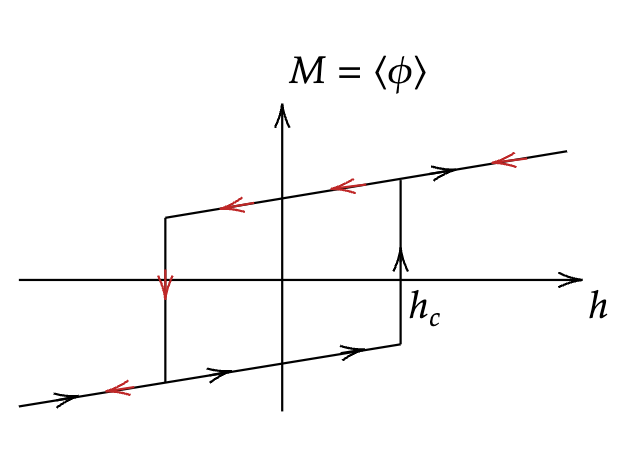}
b)\includegraphics[width=6cm]{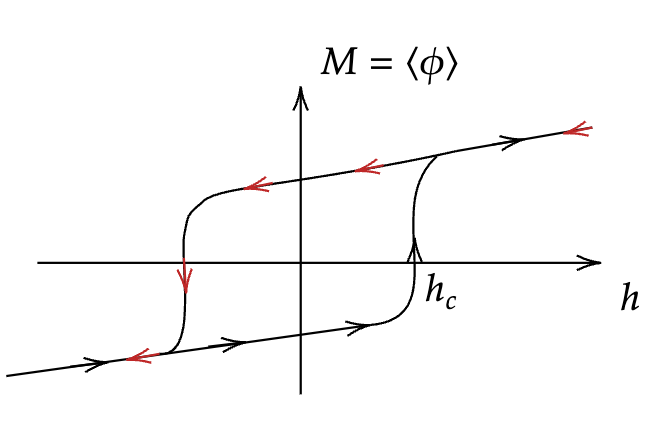}
\caption{Hysteresis loop without and with fluctuations taken into account.}
\label{Hysteresis}
\end{figure}

The Landau theory of phase transitions discussed above neglects the effect of thermal fluctuations. To take the fluctuations into account, let us first discuss the hysteresis loop.
The hysteresis loop plots spontaneous magnetization when the external magnetic field changes. 
It can be generalized to describe the phase coexistence phenomenon in generic first-order phase transitions. 
Let us first consider the mean-field theory of the hysteresis loop neglecting thermal fluctuations.
Consider the following Ising action with external magnetic field 
\be
F=a \phi^2+ \lambda \phi^4 -h \phi \cdots.
\ee
In the low-temperature phase ($a<$0), when $h\neq0$, the depth of the two vacua are different. 
Let us start with the $h\ll 0$ configuration, given in Fig.~\ref{Hpotential} a), and slowly increase $h$. 
The black dot indicates the location of the vacuum. 
As $h$ becomes slightly bigger than 0, the vacuum remains stuck in the $\phi < 0$ meta-stable vacuum, since in mean field theory approximation we have neglected fluctuations that can help the system bypass the energy barrier to the true vacuum at $\phi > 0$,
see Fig.~\ref{Hpotential} b). 
As $h$ increases further, at a critical $h_c$, the barrier between the meta-stable and the stable vacua disappears: this is when the phase transition happens. 
Fig.~\ref{Hysteresis} a) shows the hysteresis loop when thermal fluctuations are neglected. 
Notice the hysteresis loop contains sharp edges, which get smoothed out when fluctuations are taken back into account. 
This is due to the fact that fluctuations can cause the vacuum to tunnel from the meta-stable vacuum to the true vacuum. 

\begin{figure}[ht]
\includegraphics[width=8cm]{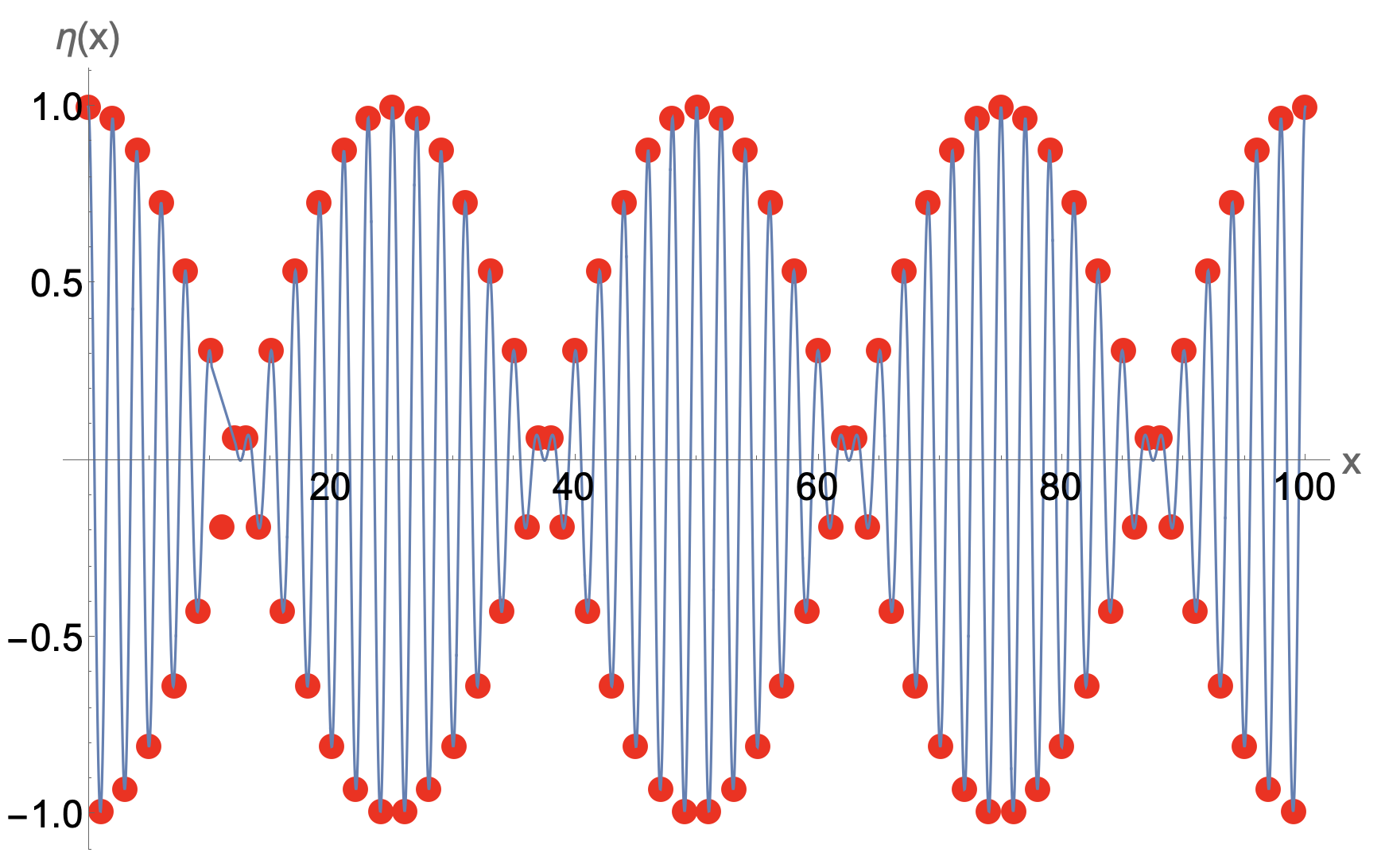}
\caption{The modulated critical mode $\eta(x)=\phi(x) \times e^{ \im k x}$, with $\phi(x)=\cos(\delta k x)$. We choose $k=\pi$ and $\delta k=0.04\pi$. The function is only evaluated on the one-dimensional lattice $x\in \mathbb{Z}$, as shown by the red dots. 
Here $k$ stands for the momentum of the critical mode, which depends on the UV details of the material. 
The $\delta k$ is the characteristic momentum of spatially modulated large-scale fluctuation. Typically, $\delta k\ll k$.}
\label{spatialmodulation}
\end{figure}

Besides changing the hysteresis loop, the thermal fluctuations can also change the critical exponents of a second-order phase transition. 
To understand this, we consider again the two-dimensional order-disorder phase transition discussed in the previous section, see Fig.~\ref{criticalline}. 
We now allow the critical mode to have spatially modulated fluctuations
\be
\eta(\vec{x})= \phi(\vec{r}) \times e^{\im \vec{k} \cdot \vec{x}} u(\Vec{r}),
\ee
with $u(\Vec{r})$ defined in \eqref{latticeDelta}. See Fig. \ref{spatialmodulation} for an illustration of the spatially modulated critical mode.
We assume the scale of the spatial modulation is much larger than the lattice scale. 
For example, taking $\phi(\Vec{x})=\phi_0 \cos(\delta \vec{k}\cdot \vec{x})$, with $|\delta \vec{k}|\ll |\vec{k}|$, the fluctuation becomes 
\be
\eta(\vec{x})\sim \phi_0 e^{\im (\vec{k}+\delta \vec{k})\cdot\vec{x}}+\phi_0 e^{\im (\vec{k}-\delta \vec{k})\cdot\vec{x}} +c. c. .
\ee
Clearly, a large-scale spatial modulation corresponds to a shift in the momentum of the order parameter. 

The Lifshitz condition (which will be explained in Section \ref{Lifshitz}) makes sure that our order parameter is stable against such a shift of momentum. 
Spatially modulated fluctuations $\phi(x)$ cost more energy compared to the homogeneous configuration. 
Therefore the effective action should contain extra terms that tend to suppress these fluctuations, such as 
$$\int dx^3 \frac{1}{2} \left(\Vec{\nabla} \phi(\Vec{x})\right)^2.$$
Since the scale of the fluctuations of $\phi(x)$ is much larger than the lattice scale, we can treat $\phi(\vec{x})$ as a continuous function in the Euclidean space ${R}^3$. 
The $\Vec{\nabla}$ is simply the spatial derivative in the continuum. 
The space group symmetry also restricts the possible derivative terms that can appear in the effective action. Take the Ising model as an example, the full Landau action is 
\be{}\label{Isingfullaction}
S= \int dx^3 \frac{1}{2} \left(\Vec{\nabla} \phi(\Vec{x})\right)^2 +a \phi(x)^2 +\lambda \phi(x)^4 +\ldots.
\ee

\begin{figure}[ht]
\includegraphics[width=8cm]{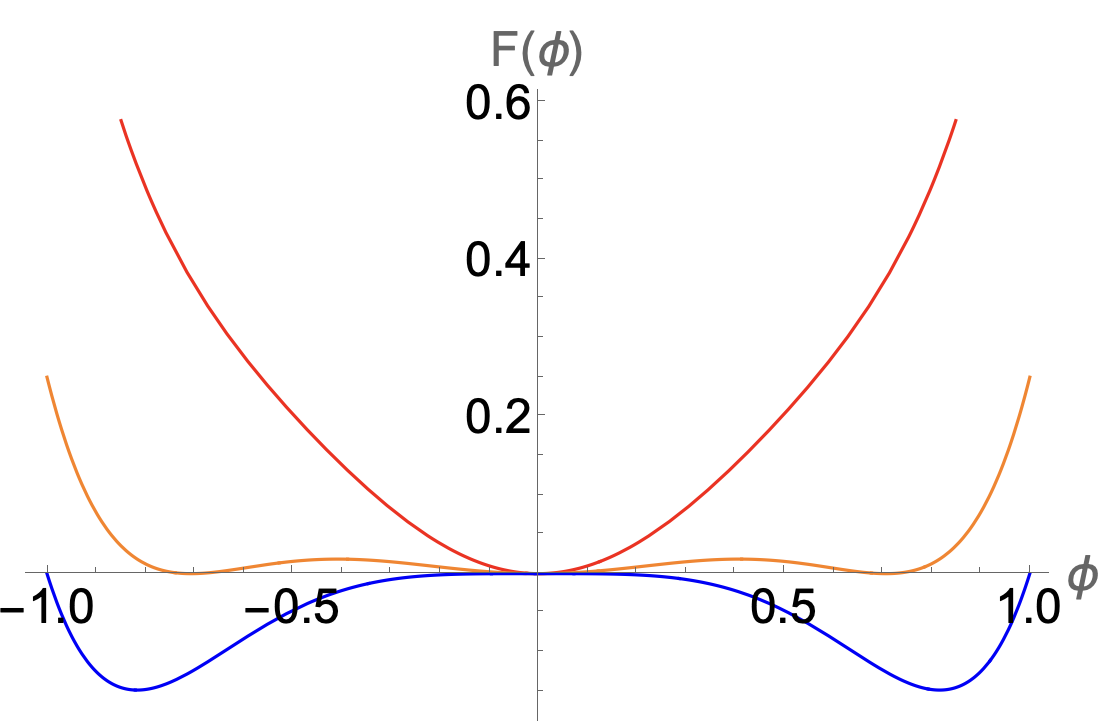}
\caption{The free energy $F(\phi)=a \phi^2-\phi^4+\phi^6$. 
The red, orange, and blue curves correspond to $a$=1, 1/4, and 0 respectively.
Notice that at $a=1/4$, there exists a free energy barrier between the $\phi=0$ state and the $\phi=\pm v$ states. This is a sign of first-order phase transitions.}
\label{phi4negative}
\end{figure}

\begin{figure}[ht]
\includegraphics[width=8cm]{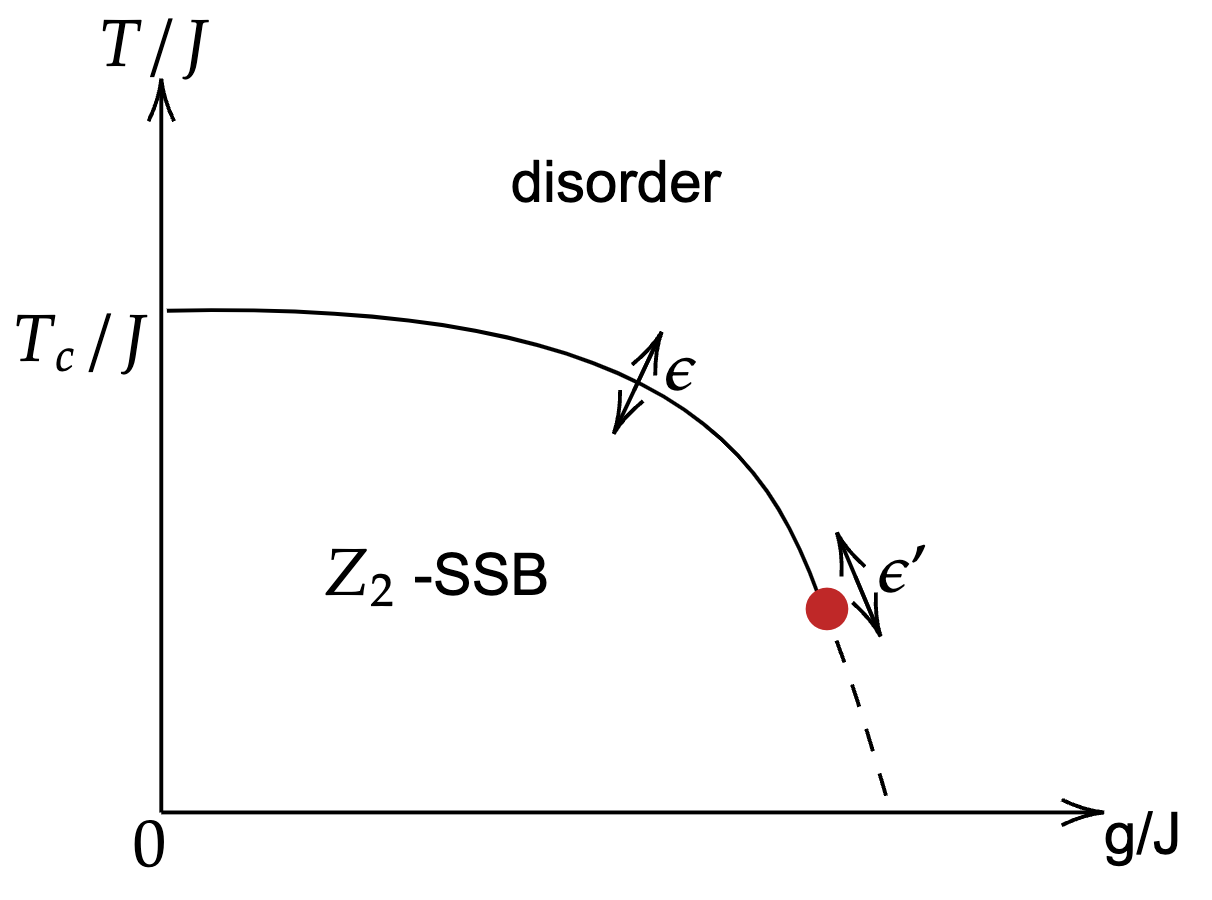}
\caption{A schematic phase diagram of the Blume-Capel model. 
The solid line corresponds to second-order phase transitions, while the dashed line is first order. 
The red dot is a tri-critical point which is given by the tri-critical Ising CFT.}\label{triIsing}
\end{figure}

Landau theory gives us the mean-field theory values of the critical exponents $\alpha=0$ and $\beta=1/2$. 
When thermal fluctuations are taken into account, the critical exponents can deviate from their mean-field theory values. 
From a modern point of view, second-order phase transitions are described by a special type of quantum field theory called conformal field theories (CFTs). 
Different universality classes correspond to different CFTs, for a review see \cite{Pelissetto:2000ek}. 
Another important concept of quantum field theory is the renormalization group. 
We will review the basic concepts of the renormalization group theory, omitting many details. 
Interested readers should refer to \cite{RevModPhys.47.773,RevModPhys.70.653,Pelissetto:2000ek}. 
Renormalization theory tells us that physics at different length scales is controlled by a set of equations called the renormalization group equations. 
Take the Ising model \eqref{Isingfullaction} as an example, the couplings constants of the action depend on the length scale $l$ through
\bea{}\label{betafunctionsRGequations}
l \frac{da}{dl}=\beta_1(a,  \lambda),\nonumber\\
l \frac{d \lambda}{dl}= \beta_2(a, \lambda).
\eea
The length scale $l$ can be understood as a cutoff scale.
In the Wilsonian picture, we coarse grain out all the microscopic physics smaller than this scale, by integrating out modes with momentum higher than $1/L$ \cite{RevModPhys.47.773}. 
The beta functions are in general complicated, and can only be calculated in certain perturbative limits \cite{PhysRevLett.28.240}. 
The RG equations have fixed points, at which the coupling constants are scale invariant. This means 
\bea{}\label{fixpointequation}
\beta_1(a,  \lambda)=\beta_2(a,  \lambda)=0.
\eea
At these fixed points, the quantum field theory invariant under the Euclidean group becomes also scale invariant. 
Usually, the Euclidean symmetry and scaling symmetry get enhanced to a bigger symmetry called the conformal group \cite{Polyakov:1970xd,POLCHINSKI1988226}. 
The corresponding quantum field theory is therefore a conformal field theory (CFT). 
The terms ``fixed point'' and ``conformal field theory'' are often used interchangeably.
Near these fixed points, one can linearize the equation by the ansatz 
$a=a_*+\delta a$ and $ \lambda= \lambda_{*}+\delta  \lambda$,
 to get
\be{}
l\frac{d}{dl}
\left(
\begin{array}{c}
 \delta a \\
 \delta  \lambda \\
\end{array}
\right)=\left(
\begin{array}{cc}
\frac{ \partial{\beta_1} }{\partial a}& \frac{ \partial{\beta_1} }{\partial  \lambda} \\
 \frac{ \partial{\beta_2} }{\partial a} & \frac{ \partial{\beta_2} }{\partial  \lambda} \\
\end{array}
\right)_{a=a_*,\lambda=\lambda_*}
\left(
\begin{array}{c}
 \delta a \\
\delta  \lambda \\
\end{array}
\right).
\ee
Here $a_*$ and $\lambda_*$ satisfies \eqref{fixpointequation}.
We care about large-scale physics, therefore the $l\rightarrow \infty$ limit.
The matrix in the above equations is sometimes called the stability matrix. 
These equations can be solved by diagonalizing the stability matrix,
\be{}
l\frac{d}{dl}
\left(
\begin{array}{c}
{\delta a'} \\
  {\delta  \lambda'} \\
\end{array}
\right)=\left(
\begin{array}{cc}
\omega_1& 0 \\
 0 & \omega_2 \\
\end{array}
\right)
\left(
\begin{array}{c}
 \delta a' \\
\delta  \lambda' \\
\end{array}
\right).
\ee
Here $\delta a'$ and $\delta  \lambda'$ are linear combinations of $\delta a$ and $\delta  \lambda$ constructed from the eigenvectors of the stability matrix. The solutions to these equations are 
\be{}
\delta a' =c_1 l ^{\omega_1} ,\quad \delta  \lambda'=c_2 l^{\omega_2} .
\ee
Clearly, as $l$ increases, coupling constants grow if $\omega>0$, while decay if $\omega<0$. 
The terms in the action whose coupling constants grow (decay) as the scale increases are called relevant (irrelevant) operators. 
The RG flow of the Ising model has two fixed points with no relevant terms. 
They are the so-called high temperature fixed point located at $a=+\infty$ and $ \lambda=0$, and the low temperature fixed point located at $a=-\infty$ and $ \lambda=0$. 
For the Ising model in three dimensions, there is another fixed point, which has only one relevant operator. 
This is the famous Wilson-Fisher fixed point \cite{PhysRevLett.28.240} describing the lattice Ising model at the critical temperature. 
The number of relevant operators at the fixed point corresponds to the number of physical parameters that we need to tune to reach the fixed point. 
The critical point at $T_c$ can be reached by tuning the temperature alone, this is because the critical Ising CFT has only one relevant operator (which preserves the $Z_2$ symmetry).
We can deform the Ising model to study tri-critical phenomena. 
We first allow the Ising spins to take value in $\sigma_i={0,\pm1}$, and then add a coupling that favors the $\sigma_i=0$ state. The Hamiltonian becomes
\be{}
H=-J\sum_{\langle ij\rangle}\sigma_i\sigma_j+ g\sum_i (\sigma_i)^2.
\ee
This is the famous Blume-Capel model. The $Z_2$ symmetry that flips all the spins are still preserved. A Monte Carlo study of this model was performed in \cite{PhysRevB.22.445}.
Figure \ref{triIsing} is a schematic phase diagram of this model.
The tri-critical point is described by the tri-critical Ising CFT. 
As is clear from the phase diagram, the tri-critical Ising CFT can only be reached by tuning two physical parameters together, the temperature and $g$. 
This is because the tri-critical Ising CFT has two relevant operators, $\epsilon$ and $\epsilon'$, which are analogous to the $\phi^2$ and  $\phi^4$ operators of free theory. 
Moving from the tri-critical Ising point to the Ising critical point on the fixed line of the phase diagram corresponds to perturbing the tri-critical Ising point by the $\lambda \epsilon'$, which triggers an RG flow towards the Ising model if $\lambda>0$. 
If $\lambda<0$, on the other hand, the phase transition becomes first order. 
This is shown schematically in Fig. \ref{phi4negative}.
This phase diagram in Fig.~\ref{triIsing} is not restricted to the Blume-Capel model, see for example \cite{PhysRevB.9.4964,herrmann1993stability}. 
The reason that the (tri-)critical Ising CFT appears in different lattice models is again deeply related to the concept of ``universality''. 
Lattice models with different ultra-violet (short distance) details, after renormalization, can flow to the same infra-red (long distance) CFT. 
Since the critical behavior of a second order phase transition is completely fixed by the corresponding CFT, their critical exponents will be the same.

\subsubsection{The Lifshitz conditions}\label{Lifshitz}
\begin{figure}[h]
\includegraphics[width=12cm]{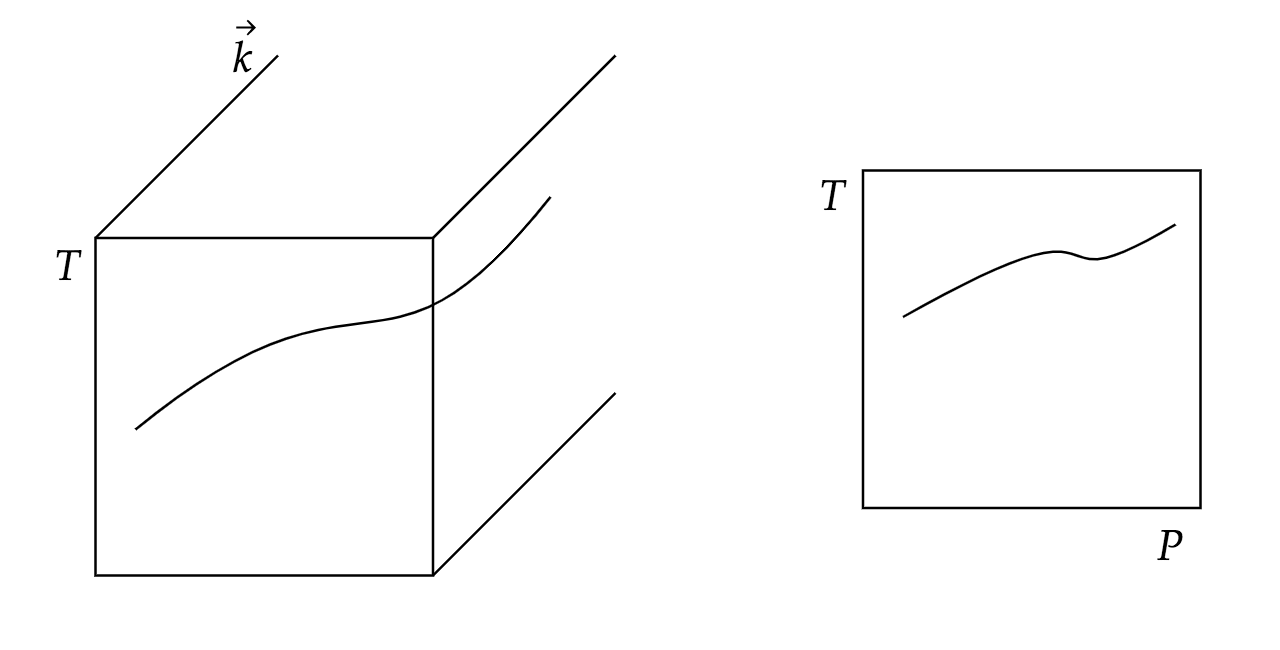}
\caption{Critical solution in the $\{T,P,\Vec{k}\}$ space and its projection on the $T$-$P$ plane. }
\label{2dtransition}
\end{figure}

In general, the mass order parameter $a(T,P)$ in the Landau theory 
\be
F= a(T,P) \sum_i \phi_i\phi_i +\ldots
\ee
depends on physical parameters, where $T$ and $P$ stands for temperature and pressure. 
Here $\phi_i$'s denote a single faithful irreducible representation of the space group, as explained in Section \ref{landautheoryandimage}.
The subscript $i$ enumerates vectors in this representation.  
Notice the temperature $T$ and the pressure $P$ do not explicitly break the space group symmetry of the material, we use them as examples of such experimental parameters.
In reality, instead of considering a single mode (irrep of space group), one has to consider the effect of other modes nearby. 
In particular, the modes with momentum close to the Brillouin zone point we consider are important. 
The mass of these modes, in addition to the physical parameters, also depends on their momentum. That is 
\be
a=a(T,P,\Vec{k}).
\ee
For the transition to be driven by the critical mode at a specific momentum, we need
\be
a(T,P,\Vec{k})=0,\quad \text{and }\quad 
\frac{\partial a(T,P,\Vec{k})}{\partial k_1}=0,\quad 
\frac{\partial a(T,P,\Vec{k})}{\partial k_2}=0, \quad
\frac{\partial a(T,P,\Vec{k})}{\partial k_3}=0.
\ee
In the five-dimensional parameter space given by $\{T,P,\Vec{k}\}$, we can at most find a one-dimensional family of solutions. 
Projection of the one-dimensional solutions onto the $(T,P)$ plane gives us a critical line so that we can reach the second-order phase transition by tuning a single parameter, see Figure \ref{2dtransition}.

We discussed the Landau condition before, which says that for the phase transition to be second order, the corresponding Landau effective action should not contain cubic terms. 
Suppose the space group of the crystal is $G$ and the order parameter that drives the phase transition lives in the 
irreducible representation $\cal R$. The Landau condition then simply means that the symmetric product of three copies of $\cal R$ should not contain the singlet representation:
\be{}
\text{ Landau condition:}\qquad {\textbf 1} \notin [{\cal R}\otimes{\cal R}\otimes{\cal R}]_{s}.
\ee
The subscript ``s'' stands for the symmetric product. 
The symbol {\textbf 1} stands for the singlet representation, in which all the group elements of $G$ act trivially. 
The number of singlet representations contained in $[{\cal R}\otimes{\cal R}\otimes{\cal R}]_{s}$ is sometimes called the Landau frequency. 
The Lifshitz condition is another necessary condition for commensurate structural phase transition to be second order. 
Let us denote the representation of $G$ that the lattice derivative $\Vec{\nabla}$ transforms in as $\mathcal{V}$, then the Lifshitz condition is 
\be{}
\text{Lifshitz condition:} \qquad \mathcal{V} \notin [{\cal R}\otimes {\cal R}]_a. 
\ee
Here the subscript ``a'' stands for the anti-symmetric product.
Notice that the lattice derivative $\Vec{\nabla}$ is invariant under spatial translations.  
The irrep $\mathcal{V}$ is also an irrep of the point group: it is the representation of the point group in which the three-dimensional vector $\Vec{r}$ transforms. 
The number of $\mathcal{V}$ representations contained in $[{\cal R}\otimes {\cal R}]_a$ is sometimes called the Lifshitz frequency. 
The Lifshitz frequency counts the number of one derivative terms allowed by the space group symmetry in the Landau effective action, such as
\be{}\label{lifshitzterm}
\int dx^3 \phi_i(x)\nabla_k {}\phi_j(x).
\ee
Notice the above term is anti-symmetric with respect to an interchange of the $i$ and $j$ index. 
This explains why the Lifshitz condition puts constraints on the anti-symmetric product representation $[{\cal R}\otimes {\cal R}]_a$.
Another way of understanding the Lifshitz condition is that it tells us that the mass of the critical mode should be located at a local minimum in the momentum space. That is 
\be{}
\frac{\partial a(T,P,\Vec{k})}{\partial k_1}=0,\quad 
\frac{\partial a(T,P,\Vec{k})}{\partial k_2}=0, \quad
\frac{\partial a(T,P,\Vec{k})}{\partial k_3}=0.
\ee
We leave the derivation to Appendix \ref{lifshitz}. 
From a renormalization group point of view, the Lifshitz terms of the form \eqref{lifshitzterm} are likely to be relevant operators for conformal field theories. 
If the space group symmetry allows such terms, without fine-tuning their couplings to zero, critical points are hard to reach. 
Clearly, the Lifshitz frequency is a property specific to the representations of the space group. 
A complete list of the Lifshitz frequency of all space groups' irreps is given in \cite{stokes1988isotropy}.

\begin{figure}[ht]
\includegraphics[width=14cm]{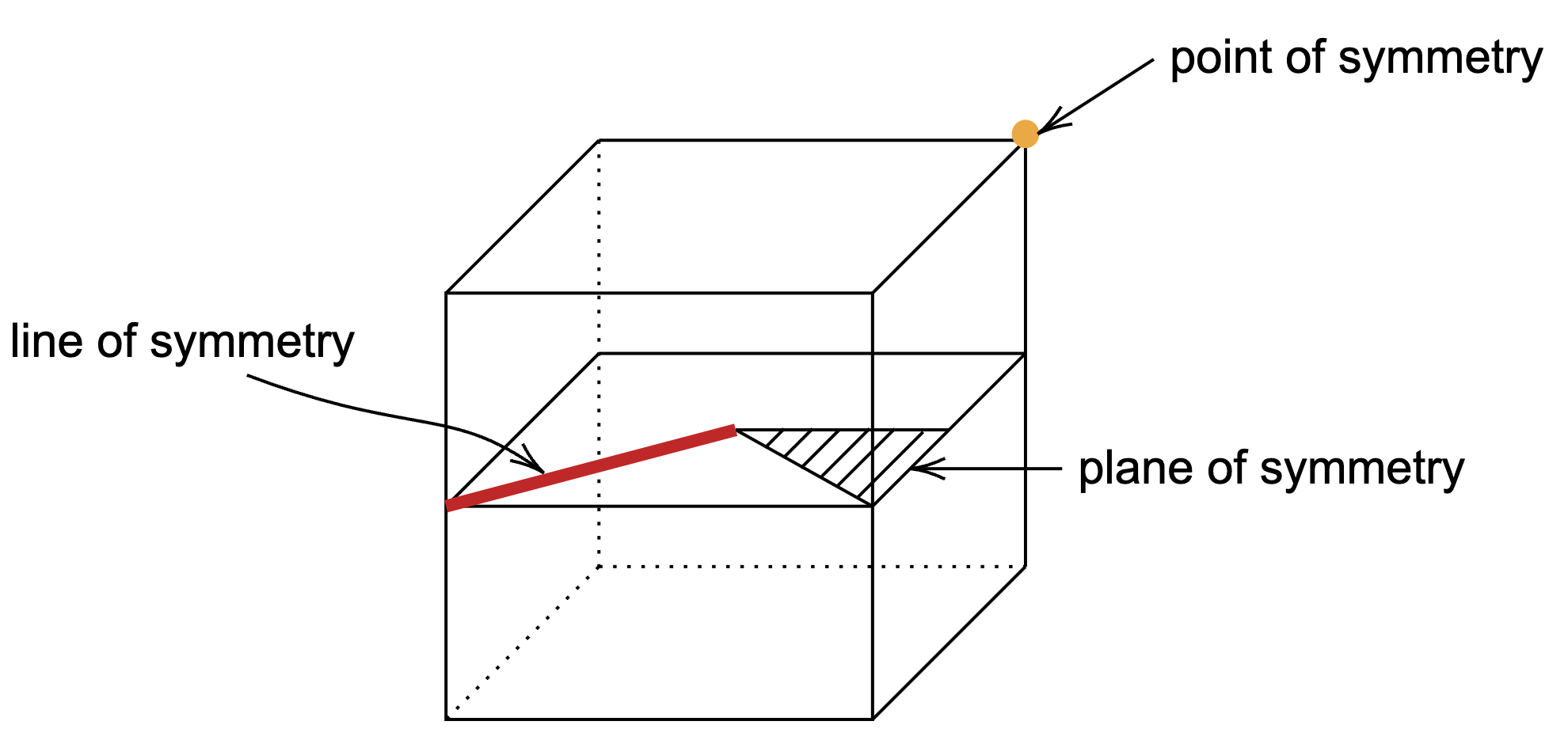}
\caption{The Brillouin zone of a square lattice. The Brillouin zone consists of points of symmetry, lines of symmetry, planes of symmetry, and generic points. }
\label{BZdomains}
\end{figure}

The points in the Brillouin zone are classified into so-called points of symmetry, lines of symmetry, planes of symmetry, and generic points. 
They denote a certain zero, one, two, and three-dimensional domain of points respectively, see Fig. \ref{BZdomains}.
For a point in the Brillouin zone, there exists a subgroup $H$ (the little group) of the space group $G$ that leaves this $\vec{k}$ point
invariant. 
Within one domain, the subgroup $H$ does not change.
It can be proven that the Lifshitz conditions are satisfied only by representations whose momentum is at ``points of symmetry'' of the  Brillouin zone.
If we allow incommensurate structural phase transitions, the order parameter driving the phase transition is not restricted to the points of symmetry. 
The incommensurate phase will have Bragg peaks whose momentum is located inside the Brillouin zone domains. 
The order parameter forms a superstructure that is incommensurate with the lattice structure.
The locations of these Bragg peaks change with temperature. 
One maybe worry about whether we should still identify incommensurate crystals at different temperatures as a single phase since different momentum points of the Brillouin zone clearly correspond to different irreps of the space group.
As explained by Michelson in \cite{michelson1978weak}, even though the momentum of the order parameter changes as temperature varies, 
the corresponding space group of this symmetry-breaking phase does not change so that they can still be identified as a single phase.  
In case of incommensurate structural phase transitions, the Lifshitz conditions can be slightly relaxed. A weaker condition called the weak Lifshitz condition was introduced by Michelson \cite{michelson1978weak}, which will be discussed in Appendix \ref{weakLifshitzsection}.

\begin{figure}[ht]
\includegraphics[width=12cm]{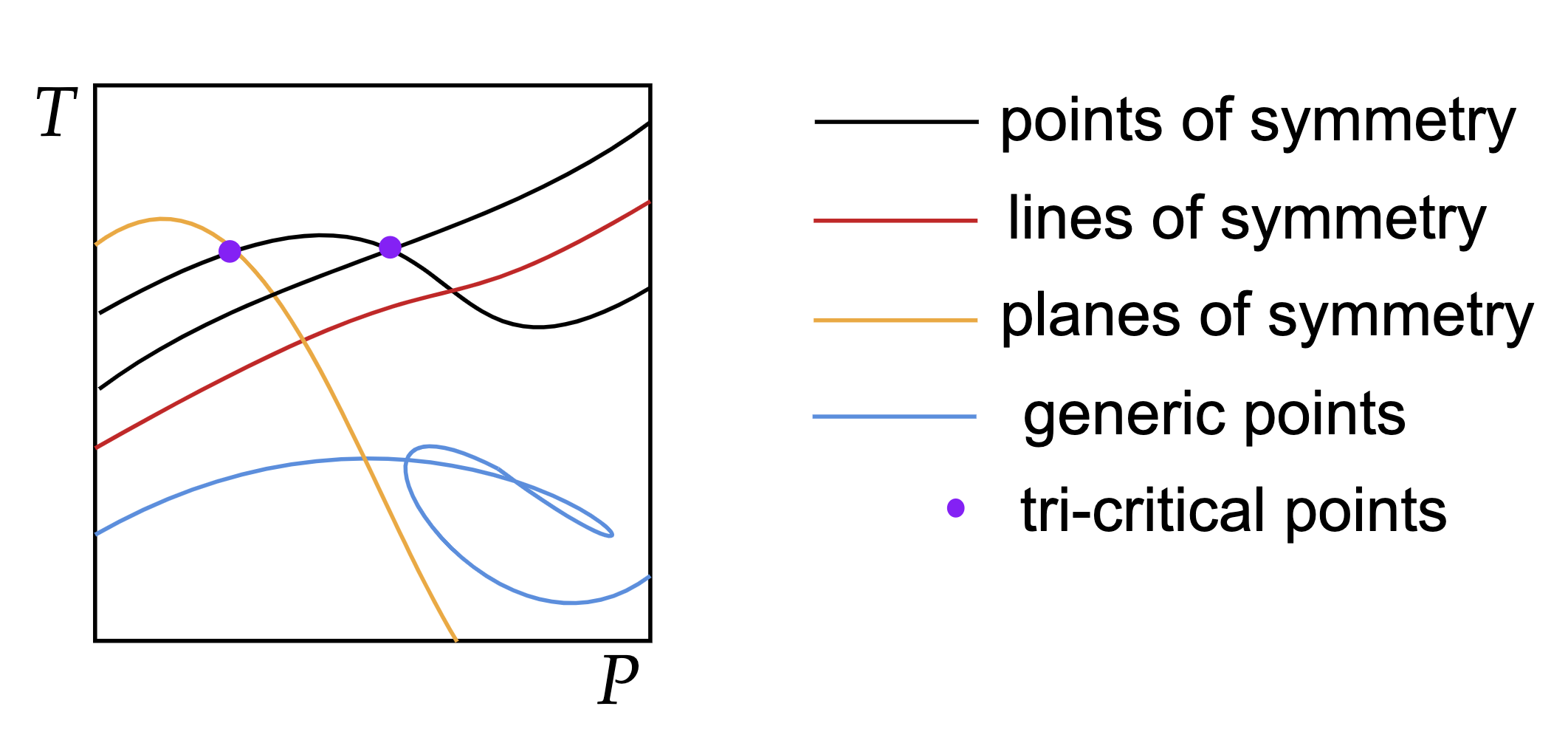}
\caption{Schematic form of the ``critical'' lines from different irreps of the space group. 
The segments at the highest temperature correspond to where the 2nd phase transition happens. 
The intersection of these pre-critical lines may give us tri-critical points which correspond to CFTs with order parameters from different irreps of the space group coupled together. }
\label{compete}
\end{figure}

In general, the irreps from points of symmetry, lines of symmetry, planes of symmetry, and even generic points may compete with each other. 
Each of these ``critical" mode correspond to a ``critical" line in the $(T,P)$ plane. 
The actual mode that drives the phase transition is the critical line at the highest temperature. 
This is shown schematically in Fig. \ref{compete}. 
The intersection of these lines may give us tri-critical points that correspond to CFTs with order parameters from different irreps of the space group coupled together. 
As far as the author is aware, no such fixed points have been observed yet in structural phase transitions. 
They are, however, natural predictions of Landau's theory. 
Recently, such tri-critical CFTs have been studied using both $4-\epsilon$ expansion and conformal bootstrap techniques \cite{Osborn:2017ucf,Rychkov:2018vya,Henriksson:2020fqi,Chai:2020zgq}.   
\subsubsection{A useful handbook and the ``ISOTROPY'' Software Suite}
The 230 crystallographic space groups have 4777 irreps whose momentum are on points of symmetry of the Brillouin zone (so that they can potentially satisfy the Lifshitz condition). 
The corresponding Landau theories, which are often called images, were classified in \cite{stokes1987images}.
Surprisingly, there are only 132 in-equivalent images. 
In the book \cite{stokes1988isotropy}, the images of all representations were listed, together with the generators of the image group $G_i$ (see Table 2), the Molien function (see Table 11), and the invariant polynomials of the corresponding Landau potential (see Table 10).
The book also contains the ``group-subgroup relations'' among the 230 crystallographic space groups. 
That is, given an irrep of a space group, the book lists all the possible subgroups (of the low-temperature phases) that this irrep can break the symmetry into. This information is given together with the Landau and Lifshitz frequencies of the irrep (See Table 1).  
Much more recently, a software suite called ``ISOTROPY''  collecting all the above was introduced \cite{stokes2007isotropy}. 
The software is also available through an online interface (\url{https://iso.byu.edu/iso/isowww.php}). 
By specifying successively the space group, the location of the momentum in the Brillouin zone, and the representation of the space group at this momentum, the software automatically generates the corresponding Landau effective action. One can also easily check the Landau frequency and the Lifshitz frequency of this representation. 
For further details, the readers should consult the user manual.

\subsection{Crystal universalities}
\label{Crystaluniversalities}\label{3dcommensurate}
As we mentioned, there are only 132 inequivalent images, which can describe commensurate structural phase transitions. 
The corresponding Landau actions are scalar field theories with $N$ scalars coupled together. 
There are images with $N=1,2,3,4,6,8,12,16$ and $26$.
We will follow the convention of \cite{stokes1988isotropy} to name the images.
Take the image ``B4a'' as an example.
The letter B tells us the number of scalars, or the dimension of the representation of the image group (A=1, B=2, C=3, D=4, E=6, F=8, G=12, H=16, J=24). 
The number $4$ in ``B4a'' is the order of the image group $G_i$.
The letter a in ``B4a'' is used to distinguish images with the same $N$ and the same order.
The image group $G_i$ of ``B4a'' is generated by the following matrix,  
\be{}
B_2=\left(
\begin{array}{cc}
 0 & 1 \\
 -1 & 0 \\
\end{array}
\right).
\ee
This group is isomorphic to the cyclic group $Z_4$. 
The Landau action of the image ``B4a'' is 
\be{}\label{B4aimage}
F(\phi)=a (\phi_1^2+\phi_2^2)+ \lambda (\phi_1^2+\phi_2^2)^2+ \lambda' (\phi_1^4+\phi_2^4) +\lambda'' (\phi_1^3\phi^2- \phi_1\phi_2^3)+\cdots.
\ee
Define $\chi=\phi_1+\im \phi_2$, then the potential can also be written as 
\be{}
F(\chi)=a \chi\chi^* +\lambda (\chi\chi^*)^2+ \lambda' \frac{1}{8}\left(\chi^4+ \chi^{*4}+6(\chi\chi^*)^2\right)-\lambda''\frac{1}{4}\im (\chi^3\chi^*-\chi\chi^{*3})+\cdots
\ee
The first two terms preserve the $O(2)$ symmetry, the third term breaks $O(2)$ to $D_4$, and the last term then breaks the $D_4$ group to $Z_4$.
For a representation, if both the Landau and the Lifshitz condition are satisfied, the representation is called ``active''. 
For a certain image, if there exists a representation satisfying both the Landau and Lifshitz conditions that map to this image, one also calls the image ``active''. 
See Table 8 of \cite{stokes1988isotropy} for images and whether they are active or not.

After knowing the images and the Landau theories, one can then study these scalar field theories in $4-\epsilon$ dimension, to analyze the renormalization group flow of these theories. 
The results are conveniently summarized in Table 12 of the book \cite{stokes1988isotropy}, where the images were listed together with whether there is a perturbatively stable fixed point. 
To be more precise, the table lists the possible subgroups of the image groups that the Landau potential (truncated to quartic order) with proper coupling constants can break the symmetry into. 
If the subgroup also lives in the attractor basin of a stable fixed point, then this phase transition is second order~\footnote{These subgroups are denoted with a double asterisk ``**'' in Table 12 of \cite{stokes1988isotropy}.
For a recent discussion about the structural transition of perovskites, in particular, the role of attractor basin of CFT fixed points, see \cite{Aharony:2022ajv}.}, at least perturbatively.
These are the universality classes we can get from structural phase transitions.

To have a better understanding of the results, we will review how they was obtained. 
It turns out that there are no active $N>8$ images \cite{stokes1988isotropy}. 
This means that the corresponding $N>8$ Landau theories can never give us to a second-order structural phase transition. 
We will focus on $N\leq8$ images. 

For images with $N=8$, only four of the images are active.
The corresponding $4-\epsilon$ fixed points with $N=8$ were studied in \cite{stokes1987continuous}, and it turns out that none of these fixed points are RG stable. 
This means that the corresponding $N=8$ Landau theories cannot be second-order structural phase transitions either. 

For $N\leq 6$ images, we start to have stable fixed points that can be realized in structural phase transitions. 
We list them in Table \ref{universality}.

\subsubsection{Perturbative fixed points}
We will now explain the perturbative fixed points in Table \ref{universality}.
\begin{itemize}
    \item The fixed point No. 1 is the Ising CFT, which is the Wilson-Fisher fixed point of 
\be{}
\mathcal{L}=\frac{1}{2}(\partial{\phi})^2+\lambda \phi^4.
\ee
The global symmetry of the CFT is the $Z_2$ group. See also Chapter 3 of \cite{Pelissetto:2000ek}.
\item The fixed point No. 2 is the XY model fixed point, which is the Wilson-Fisher fixed point of the following Lagrangian
\be
\mathcal{L}=\frac{1}{2}\sum_{i=1}^{2}(\partial\phi_i)^2+\lambda(\phi_1^2+\phi_2^2)^2.
\ee
The global symmetry of the CFT is the $O(2)$ group. See also Chapter 4 of \cite{Pelissetto:2000ek}.
\item The fixed point No. 3 is the stable fixed point of the following $N=3$ Cubic model,
\be
\mathcal{L}=\frac{1}{2}\sum_{i=1}^{3}(\partial\phi_i)^2+ u(\phi_1^2+\phi_2^2+\phi_3^2)^2+v (\phi_1^4+\phi_2^4++\phi_3^4).
\ee
The global symmetry of the CFT is the Cubic group $(Z_2)^3\rtimes S_3$. 
The three $Z_2$ symmetries flips the sign of $\phi_1$, $\phi_2$ and $\phi_3$ respectively. 
The permutation group $S_3$, on the other hand, permutates the three scalar fields.
The 1-loop renormalization group flow equations of the above Lagrangian have four fixed points, corresponding to zero solutions of the beta functions of the two coupling constants,
\be
l\frac{d}{dl} u =\beta_{u}(u,v)=0,\quad \frac{d}{dl} v =\beta_{v}(u,v)=0.
\ee
Among them, only one is stable, that is the CFT which has only one relevant operator $\sum_i \phi^i \phi^i$. In other words, one has to make sure that the stability matrix 
\be
\left(
\begin{array}{cc}
 \frac{\partial\beta_u}{\partial u} &  \frac{\partial\beta_u}{\partial v} \\
  \frac{\partial\beta_v}{\partial u} &  \frac{\partial\beta_v}{\partial v} \\
\end{array}
\right)
\ee
has no positive eigenvalues, so there is no relevant operator coming from the $\phi^4$ terms.
See also Chapter 11.3 of \cite{Pelissetto:2000ek}.
\item 
The fixed point No. 4 is a fixed point given by two copies of XY models coupled together, which we will denote as ``XY$^2$'' fixed point. 
It is the stable fixed point of the following Lagrangian
\be\label{XYsquare}
\mathcal{L}=\frac{1}{2}\sum_{i=1}^{4}(\partial\phi_i)^2+ u(\sum_{i=1}^4 \phi_i^2)^2 +v \left((\phi_1^2+\phi_2^2)^2+(\phi_3^2+\phi_4^2)^2\right).
\ee
The global symmetry of the CFT is the group $O(2)^2\rtimes Z_2$. 
The first copy of the $O(2)$ group rotates $(\phi_1,\phi_2)$, while the seconds $O(2)$ rotates $(\phi_3,\phi_4)$. The $Z_2$ symmetry, on the other hand, does
\be
\phi_1\leftrightarrow\phi_3,\quad \phi_2\leftrightarrow\phi_4.
\ee
Similarly, the RG flow equations of the above Lagrangian have four fixed points. 
At the critical temperature, only the stable fixed point is realized.

\item The fixed point No. 5 is the $N=4$ Cubic fixed point, which is also the stable fixed point of
\be
\mathcal{L}=\frac{1}{2}\sum_{i=1}^{4}(\partial\phi_i)^2+ u(\sum_{i=1}^4 \phi_i^2)^2+v (\sum_{i=1}^4 \phi_i^4).
\ee
The global symmetry of the CFT is the Cubic group $(Z_2)^4\rtimes S_4$. 
\item The fixed point No. 6 is given by three copies of XY models coupled together, which we will denote as ``XY$^3$''. 
It is the stable fixed point of the following Lagrangian
\be
\mathcal{L}=\frac{1}{2}\sum_{i=1}^{6}(\partial\phi_i)^2+ u(\sum_{i=1}^6 \phi_i^2)^2 +v \left((\phi_1^2+\phi_2^2)^2+(\phi_3^2+\phi_4^2)^2+(\phi_5^2+\phi_6^2)^2\right).
\ee
The global symmetry of the CFT is the group $O(2)^3\times S_3$. 
\end{itemize}

\subsubsection{N=6}
We now come back to the perturbative RG analysis of the images. 
The images with $N=6$ were studied in \cite{hatch1986renormalization} up to 1-loop order in the $4-\epsilon$ expansion.
A careful analysis of the results shows that the only stable fixed point is the XY$^3$ model.

\subsubsection{N$\leq$4}
In \cite{michel1981landau}, all irreducible subgroups of O(4) were classified. 
Irreducible subgroups are groups under which the four-dimensional vector representation of O(4) remains irreducible. 
This also means that the corresponding Landau theory has only one quadratic mass term. 
This constraint is related to the requirement that the potential second-order transition can be reached by tuning a single physical parameter. 
In other words, if the Landau action has more than one quadratic polynomial, all of them need to be tuned to zero to reach criticality. The corresponding CFTs can at most be tri-critical points.

Based on this result, the paper \cite{toledano1985renormalization} studied the perturbative RG flow of these Landau theories.
\begin{table}[h]
\begin{tabular}{|l|l|}
\hline
number of scalars & fixed points                                                          \\ \hline
1                 & Ising                                                                 \\ \hline
2                 & XY                                                                    \\ \hline
3                 & the O(3) vector model, the $N$=3 Cubic model                          \\ \hline
4                 & the O(4) vector model, XY$^2$, the $N$=4 Cubic model, the (hyper)tetrahedral \\ \hline
\end{tabular}
\caption{The stable perturbative fixed point with up four scalars. (Tri-critical fixed points with two mass terms are not considered.)}\label{perfixepoints1234}
\end{table}
The RG flows were studied up to two-loop order. 
The symmetry group $H$ of the RG flow (or the lattice model) can be either smaller or equal to the symmetry group $G$ of the fixed point. 
We call a fixed point with symmetry group $G$ stable, as long as it is stable against all perturbations that respect the symmetry group $G$. 
It is of course also possible for the fixed point to be further stable against perturbations that respect a smaller group $H$ when the ``emergent symmetry'' phenomenon happens. 
It was found in \cite{toledano1985renormalization} that there are only four stable fixed points. They are the 
\be{}
\text{XY}^2,\quad \text{$N$=4 Cubic}
\ee
fixed points we already discussed, and two extra fixed points that will not be realized in structural phase transitions.
\begin{itemize}
    \item The first extra fixed point is the tetrahedron fixed point of the Lagrangian 
\be{}
\mathcal{L}=\frac{1}{2}\sum_{i=1}^{4}(\partial\phi_i)^2+ u(\sum_{i=1}^4 \phi_i^2)^2 +v \sum_{ijklm} d_{ijm}d_{klm}\phi_i\phi_j\phi_k\phi_l.
\ee
The model was introduced in \cite{Zia:1975ha}.
The invariant tensor $d_{ijk}$ is an invariant tensor of the $S_5$ group, which can be calculated using the procedure discussed in \cite{Zia:1975ha}. 
The global symmetry of this CFT is the group $S_5\times Z_2$. 
This is the symmetry group of a (hyper)tetrahedron with five vertices, which can be embedded in the four-dimensional Euclidean space.
In group theory language, the symmetric group $S_5$ has a four-dimensional irreducible representation, sometimes called the "standard" representation.
We denote the CFT ``the (hyper)tetrahedral" in Table \ref{perfixepoints1234}.
\item 
The second extra fixed point is the $O(4)$ vector model 
\be
\mathcal{L}=\frac{1}{2}\sum_{i=1}^{4}(\partial\phi_i)^2+ \lambda (\sum_{i=1}^4 \phi_i^2)^2.
\ee
The global symmetry of the CFT is clearly $O(4)$.
\end{itemize}
In the paper \cite{michel1981landau}, the possible Landau theories with N$\leq$ 3 scalar fields were also listed, again based on knowledge of irreducible subgroups of O(2) and O(3).
One can similarly work out the possible irreducible fixed points, which are also listed in Table~\ref{perfixepoints1234}. Recently, we generalized the work of~\cite{michel1981landau} to the five scalar models in~\cite{Rong:2023xhz}.

Notice even though the main topic of our paper is structural phase transitions, the classification in \cite{michel1981landau,hatch1986renormalization} applies to all phase transitions which can be described by up to four scalars coupled together. 
Many of the irreducible subgroups and Landau theories may seem purely theoretical at the beginning, but they later do appear in interesting condensed matter systems. 
As an example, the group $GL(2,\mathbb{Z}_3)$ (or $[D_3/C_2;O/D_2]$ in the convention of \cite{michel1981landau,hatch1986renormalization}) is the symmetry group of the effective action that describes a certain frustrated Ising model on the Kagome lattice \cite{Huh:2011pp}.

\subsubsection{Comments on non-perturbative results}
Notice that the renormalization group flow analysis we discussed above is perturbative in $\epsilon=4-D$.
The non-perturbative RG flow in three dimensions ($\epsilon=1$) can be different in many aspects. 
For example, perturbatively stable fixed points can become unstable and vice versa. 
The attractor basin of the stable fixed points may change. 
In general, non-perturbative physics is hard to attack. 
However, certain methods such as Monte Carlo simulation (see for example \cite{landau2021guide}) and conformal bootstrap \cite{Poland:2018epd} have greatly improved our understanding of non-perturbative RG flow.

Let us take the XY$^2$ model as an example. 
For an RG space with $ O(2)^2\rtimes Z_2$ symmetry, at two-loop order, the perturbatively stable fixed point is the fixed point where two O(2) vector models are interactively coupled. 
There exists another fixed point with 
\be
u=0,\quad v\neq 0
\ee
in \eqref{XYsquare}. Notice since $u=0$, the model can be written as two copies of XY models decoupled, one involves the scalars $\phi_1$ and $\phi_2$, another involves $\phi_3$ and $\phi_4$.
Even though this decoupled fixed point is perturbatively unstable, the non-perturbative results from both Monte Carlo simulation and conformal bootstrap suggest that it is non-perturbatively stable.
Let us denote $\epsilon_1\sim(\phi_1^2+\phi_2^2)$ and $\epsilon_2\sim(\phi_3^2+\phi_4^2)$ as the mass operators of the O(2) vector models. 
The $\epsilon$ operator of the O(2) model has scaling dimension $\Delta_{\epsilon}=1.51124(22)$ \cite{Campostrini:2006ms}. 
Since the two copies of the O(2) models are decoupled, the operator $O=I_{23}=\epsilon_1 \epsilon_2$ will not be re-normalized. 
We get  
$\Delta_O=2\Delta_{\epsilon}>3$. 
This means that the decoupled fixed point is stable when perturbed by this operator. 
In fact, all operators preserving the $O(2)^2\rtimes Z_2$ symmetry of the decoupled O(2) models are irrelevant (except for the mass operator $\epsilon_1+\epsilon_2$ whose coupling will be tuned to zero at the critical temperature). 
The decoupled O(2) fixed point is non-perturbatively stable. 
Similarly, for the XY$^3$ model, the true stable fixed point is the three copies of decoupled XY CFTs.

The Ising and XY fixed points were known to be non-perturbatively stable for a long time. 
The non-perturbative stability of the $N$=3 and $N$=4 Cubic fixed points are both non-perturbatively stable. We will discuss this in Section~\ref{QLMandCubicCFT}.

Images with the same universality class can have different critical exponent $\omega$, which controls the leading correction to the critical behavior \cite{Pelissetto:2000ek}. 
This critical exponent is related to the scaling dimension of the leading irrelevant operator allowed by the image group $G_i$.
The critical fixed point can have an enhanced symmetry, which is bigger than the symmetry group of the RG flow, the image symmetry group $G_i$.
Take the B4a image \eqref{B4aimage} as an example, for which the image group is $Z_4$. 
At the critical point, the $Z_4$ symmetry is enhanced to $O(2)$.
The quadratic polynomials allowed by $Z_4$ are 
\be
O_1=(\chi \chi^{*})^2,\quad  O_2=Re[\chi^4], \quad \text{and}\quad   O_3=\im Im[\chi^4]. 
\ee
The operator $O_1$ has scaling dimension $\Delta_{O_1}=3.794(8)$, while the operators $O_2$ and $O_3$ are the leading charge-4 operator in the spectra, with scaling dimension $\Delta_{O_2}=\Delta_{O_3}=3.11535(73)$ \cite{Chester:2019ifh}. 
A nice review of the conformal data of the O(N) vector models is \cite{Henriksson:2022rnm}.
Since the leading irrelevant operators allowed by the image group $Z_4$ are $O_2$ and $O_3$, the ``B4a'' image has $\omega=0.11535(73)$.
The image group of ``B6a'' (whose image group is $Z_6$), on the other hand, only allows the $O_1$ deformation, which has $\omega=0.794(8)$.
Many of the universalities listed in Table \ref{universality} have small $\omega$. 
This makes this corresponding second-order phase transition ``dirty''. 
For Monte Carlo simulation, this means that the finite size effects are hard to get rid of. 
For real experimental measurements, this means that one has to be very close to the critical temperature to observe a good scaling behavior of the physical quantities.

\subsection{Future Directions}
{\it Incommensurate structure phase transitions universalities.} We discussed briefly in Section \ref{Lifshitz} and Appendix \ref{weakLifshitzsection} the incommensurate structural phase transitions.
The order parameter of the symmetry-breaking phase is incommensurate with the lattice structure. 
This has been observed in materials such as  Rb$_2$ZnCl$_4$ \cite{PhysRevB.54.3115}, and the transition was shown to be in the three-dimensional XY model universality class. 
The irreps of the 230 crystallographic space groups satisfying the weak Lifshitz condition and also the Landau condition are classified in \cite{stokes1993landau}. 
It will be interesting to write down the corresponding Landau effective action and then use the knowledge of three-dimensional conformal field theory to determine the order of the phase transition, and which universality class the phase transition is in. 
A full list of possible CFTs that can be realized in incommensurate transitions will be interesting, which we leave for future work.

{\it Two-dimensional structural phase transitions.} 
We briefly mentioned the two-dimensional structural phase transitions of absorbed monolayers in Section \ref{landautheoryandimage}. 
This type of phase transition is controlled by the spontaneous breaking of two dimension space groups, which are also called wallpaper groups. There are only 17 wallpaper groups.
Classifying the irreps and analyzing the corresponding Landau theory is less laborious than in three dimensions. 
The order-disorder type of phase transitions was classified in \cite{domany1978classification,domany1979classification}. 
The irreps of the wallpaper at points of symmetry, their Landau and Lifshitz frequencies, and the subgroups that are preserved by these irreps were later worked out in \cite{hatch1984symmetry}, which therefore include a classification of possible displacive type transitions. 
The effective actions of all irreps of the wallpaper group satisfying both Landau and Lifshitz conditions were worked out in \cite{rottman1981symmetry}.
The effective actions appearing in these studies include the actions of the two-dimensional Ising model, the three-state Potts model, the four-state Potts model, the clock model, the three-state Cubic model and etc.   
One should, however, be careful when using the Landau and Lifshitz conditions to infer the order of the transitions because of the strong thermal fluctuations.
Typical second-order phase transitions which violate the Landau condition include the three-state and four-state Potts models.
We believe the (weak) Lifshitz condition should also be used with caution.
It will be interesting to study more carefully these two dimensional Landau actions more using the knowledge of two-dimensional CFTs in the future.

{\it Magnetic Phase Transitions.} Table \ref{universality} lists the conformal field theories that can be realized in commensurate structural phase transitions.
One can potentially do a similar analysis for magnetic transitions. This type of phase transition can be analyzed using the representation theory of the so-called ``magnetic space group (Shubnikov groups)''~\cite{toledano1987landau}, which describes magnetic phase transitions through group-subgroup relations just like the crystallographic space group that describes structural phase transition, see for example \cite{birman1980tensor}. 
A full list of possible CFTs that can be realized in magnetic phase transitions will be an interesting problem to attack in the future.

\section{How to classify all the phases of a Landau theory using GAP?}\label{phasesofLandau}
As we have seen in the previous section, the effective theory of structural phase transition sometimes gives us Landau actions with complicated symmetry groups.
In this section, we discuss how to classify all the phases of a Landau theory using GAP~\cite{GAP4}. 
We will assume that the readers know the basics of coding in GAP. 
Example codes are provided in the attached files. 
Nowadays, the best way to learn GAP is to ask questions in ChatGPT~\cite{openai2025chatgpt}. 
In this section, we will follow the convention in GAP and denote the permutation group and cyclic group as SN and CN respectively.

\subsection{Example 1: The Cubic model}\label{CubicLandau}
The Cubic group is generated by 
\begin{align}
    Z=\left(
\begin{array}{ccc}
 -1 & 0 & 0 \\
 0 & 1 & 0 \\
 0 & 0 & 1 \\
\end{array}
\right),\quad g_1=\left(
\begin{array}{ccc}
 0 & 1 & 0 \\
 1 & 0 & 0 \\
 0 & 0 & 1 \\
\end{array}
\right),\quad {\rm and }\quad  g_2= \left(
\begin{array}{ccc}
 0 & 1 & 0 \\
 0 & 0 & 1 \\
 1 & 0 & 0 \\
\end{array}
\right)
\end{align}
In this paper, we denote the symmetry of the Landau action as ``Sym''. In our case, Sym=Cubic=S3$\ltimes ({\rm C2})^3$.
The Molien function is 
\begin{align}
    M(z)=\frac{1}{(1-z^2)(1-z^4)(1-z^6)}.
\end{align}
This indicated that there are three basic invariant polynomials, see Section~\ref{Landautheory}, which are 
\begin{align}
    &I_2(\phi)= (\phi^1)^2+(\phi^2)^2+(\phi^3)^2,\nonumber\\
    &I_4(\phi)= (\phi^1)^4+(\phi^2)^4+(\phi^3)^4,\nonumber\\
    &I_6(\phi)= (\phi^1)^2(\phi^2)^2(\phi^3)^2.
\end{align}
The generic Landau action can be written as 
\begin{align}
    V(\phi)= \sum_{n=1}^{\infty} a_{n,2} \big( I_2(\phi) \big)^n +\sum_{n=1}^{\infty} b_{n,4} \big( I_4(\phi) \big)^n+ \sum_{n=1}^{\infty} c_{n,6} \big( I_6(\phi) \big)^n.
\end{align}
We now want to understand all possible phases of the action. 
The details of the algorithm are summarized in \cite{kim1986classification}. 
Unfortunately, this algorithm has been largely forgotten by physicists of the current generation. We now review a few group theory basics.

If a subgroup $H$ of group $G$ is to be a symmetry group of the lower symmetry phase after a transition, the irrep $\Gamma$ of $G$ that the order parameter transforms in must subduce the identity representation of $H$ at least once~\cite{kim1986classification}.
The number of times the identity representation appears is called the subduction frequency or subduction index. 
This is the subduction criterion \cite{PhysRevLett.17.1216}.
The group $H$ is then
called an ``isotropy subgroup" of the representation $\Gamma$.
The physical meaning of the subduction frequency is the number of independent vectors invariant under the isotropy group.
If the isotropy subgroup $H_1$ and $H_2$, with $H_1\subset H_2$, have the same subduction frequency, then $H_1$ can not be a symmetry broken phase. 
Classifying the phase of a Landau action is then just to classify all isotropy subgroups of $\Gamma$.

What is the degeneracy of a phase? The degeneracy of a phase is given by $|G|/|H|$, see \cite{RevModPhys.52.617}.

The problem of classifying the isotropy subgroups of a theory is usually tedious. Nowadays, we can however use the computer algebra system GAP \cite{GAP4} to speed up the calculation greatly. The GAP system has been used in physics quite extensively, in particular, in the classification of perturbative CFTs (universality classes) in 4-$\epsilon$ dimensions~\cite{Rong:2023xhz}.

To classify the phases, we first calculate all the subgroups up to conjugation, which can be easily done using the function ``ConjugacyClassesSubgroups()''. Notice that the two phases which respect two conjugate subgroups are equivalent.
We then calculate the subduction frequency by the following formula
\be
n^R=\frac{1}{|G|}\sum_{C\in {\rm \{conj.\; classes\}}}N_{C}\bar{\chi}^{\rm Id}(C)\chi^R(C).
\ee
Here $N_C$ is the number of group elements in a specific conjugacy class, and $\chi^R(C)$ is the character of the representation. The character of the identity representation is given by $\chi^{\rm Id}(C)=(1,\ldots,1)$. One can easily calculate the above inner product by the GAP function ``ScalarProduct()''.

\begin{figure}
\centering
\includegraphics[scale=0.5]{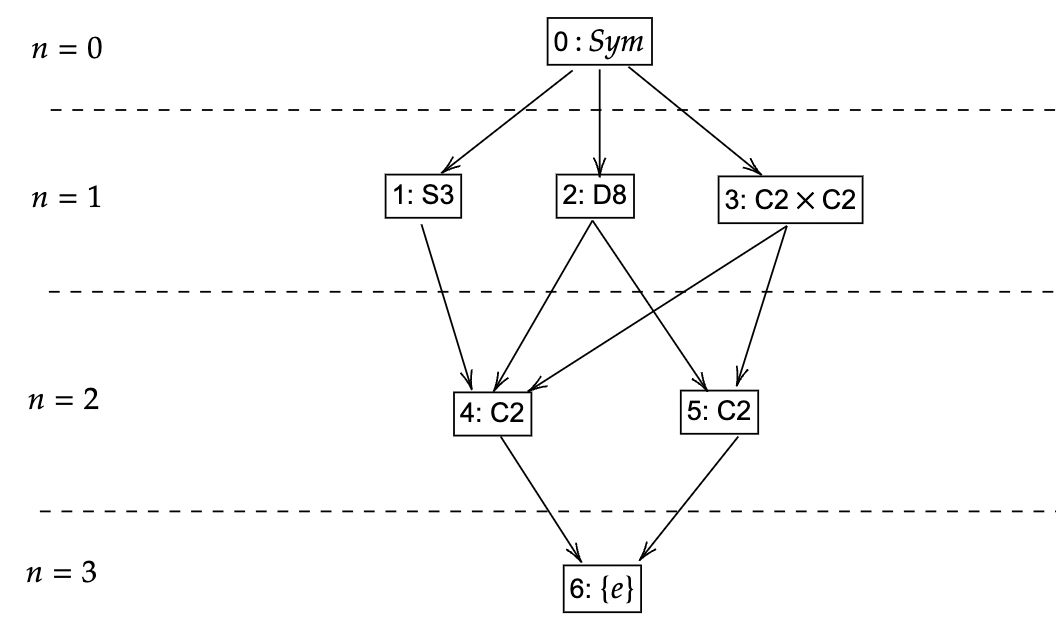}
\caption{The group-subgroup relations between phases of the Cubic model.}\label{relationsphases}
\end{figure}

By successively calculating the subgroups of subgroups, and calculating the subduction index, we can classify all possible phases. The result is summarized in Table~\ref{phasesofCubic}.
\begin{table}[h]
\begin{tabular}{|l|l|l|l|l|l|}
\hline
No. & group        & inv. vector & generators              & name              & sub. ind. \\ \hline
0   & Sym          & $(0,0,0)$   & $Z,~g_1,~g_2$             & disorder          & 0         \\ \hline
1   & S3           & $(a,a,a)$   & $g_1,~g_2$               & corner cubic      & 1         \\ \hline
2   & D8           & $(a,0,0)$   & $Z.g_1,~Z$              & face-center cubic & 1         \\ \hline
3   & C2$\times$C2 & $(a,a,0)$   & $g_1,~(g_2).Z.(g_2)^{-1}$ & bond-center cubic & 1         \\ \hline
4   & C2           & $(a,a,b)$   & $g_1$                   &                   & 2         \\ \hline
5   & C2           & $(0,a,b)$   & $Z$                     &                   & 2         \\ \hline
6   & $\{e\}$      & $(a,b,c)$   &                         &                   & 3         \\ \hline
\end{tabular}
\caption{The phases of the Cubic model.}\label{phasesofCubic}
\end{table}
Their group-subgroup relations are summarized in Fig.~\ref{relationsphases}. We do not include relations that can be decomposed into successive group-subgroup relations.
We provide the code for this calculation in the attached file ``GAP$\_$cubic.g''.

\begin{figure}[h]
\centering
a)\includegraphics[scale=0.5]{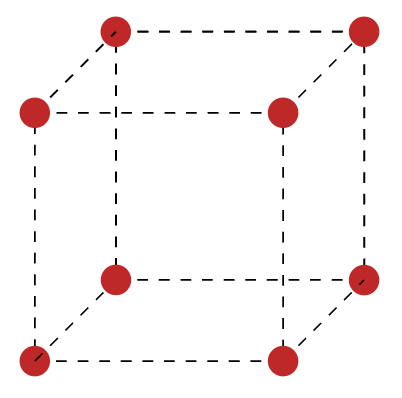}
b)\includegraphics[scale=0.5]{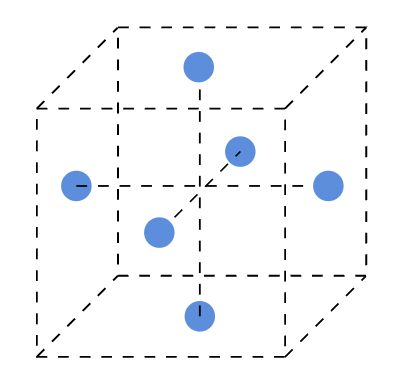}
c)\includegraphics[scale=0.5]{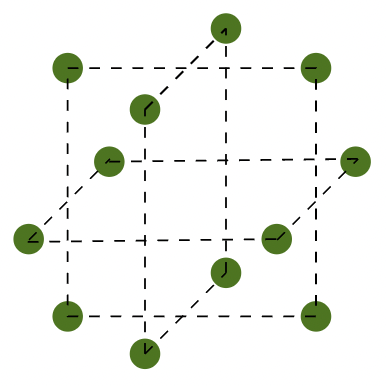}
\caption{a) The corner cubic phase. b) The face-center cubic phase. c) The bond-center cubic phase.}\label{cubephases}
\end{figure}

The maximal phases (those with subduction index 1), are the corner cubic, face-center cubic, and bond-center cubic phases. The names come from the location of the invariant vector at the surface of a cube, as summarized in Fig.~\ref{cubephases}.
The point $(\phi^1,\phi^2,\phi^3)=(1,1,1)$ is in corner cubic phase corresponds, which lead to 
\begin{align}
    I_1=3,~I_2=3,~{\rm and}~ I_3=1.
\end{align}
We can therefore reverse engineer a Landau potential whose minimus leads to the corner cubic phase, which is 
\begin{align}
    V(\phi)=(I_2(\phi)-3)^2+(I_4(\phi)-3)^2+(I_6(\phi)-1)^2.
\end{align}
We can similarly reverse-engineer the Landau potential which favors the other phases.

The above trick can be generalized to other finite groups. Given a vector, it belongs to an orbit of the group. 
For compact Lie groups and finite groups, the invariant polynomial separate orbits.
For finite groups, an orbit (or a phase) is completely specified by $l$ invariant polynomials~\cite{kim1986classification}. 
Here $l$ is the dimension of the order parameter, in the case of the Cubic model, we have $l=3$.
This also helps us to engineer a potential that realizes a specific phase. 
Suppose an orbit is specified by the condition $P_i(\phi)=c_i$. Then the potential 
$$V(\phi)=\sum_i (P_i(\phi)-c_i)^2,$$
will have minimums realizing the phase.

How do we construct the invariant polynomial? As we have mentioned previously, the Molien Series counts the number of invariant polynomials at a specific degree, which can be calculated in GAP using ``MolienSeries()''. To explicitly construct the invariant polynomial, one can use the so-called Reynolds operator. One can pick a random polynomial $f(\phi_i)$, the Reynolds operator is just
\be
R(f)=\frac{1}{|G|}\sum_{g\in G} f\bigg (\rho(g)_i^j\phi_j\bigg).
\ee
This means that we need to act on the initial polynomial by the group elements, and then average over the group.

Phase indicators for Monte Carlo simulation: In the thermal dynamical limit, phases connected by the action of the group Sym are different super-selection sectors. 
The probability of transition from one sector to the other is almost surely zero. 
When we do Monte Carlo simulation on finite-sized lattice, however, the transition probability is non-zero.
It will be nice to have phase indicators that are invariant under the action of Sym.  
These will be the invariant polynomial, or more precisely, the invariant harmonics on unit $S^3$, that is the Binder ratios 
\be
B^{(d)}_i=\frac{\langle P^{(d)}_i(\phi)\rangle}{\langle\phi^2\rangle^{d/2}}.
\ee 
Here $P^{(d)}_i(\phi)$ are the invariant polynomials. 
We will discuss the renormalization of the Cubic model in Section \eqref{QLMandCubicCFT}.

\subsection{Example 2: A dimer model on a Kagom\'e lattice}
We now consider a more complicated Landau action, which was discovered in~\cite{huh2011vison}. It corresponds to an effective action that describes the quantum dimer model on Kagom\'e lattice. 
Readers who are interested in physics should read~\cite{huh2011vison}, we will only discuss the mathematical problem of finding all phases of the potential. 
The group that leaves the Landau action is generated by 
\bea
&&T_u=\left(
\begin{array}{cccccccc}
 0 & 0 & -1 & 0 & 0 & 0 & 0 & 0 \\
 0 & 0 & 0 & -1 & 0 & 0 & 0 & 0 \\
 e^{\frac{2 i \pi }{3}} & 0 & 0 & 0 & 0 & 0 & 0 & 0 \\
 0 & e^{\frac{2 i \pi }{3}} & 0 & 0 & 0 & 0 & 0 & 0 \\
 0 & 0 & 0 & 0 & 0 & 0 & -1 & 0 \\
 0 & 0 & 0 & 0 & 0 & 0 & 0 & -1 \\
 0 & 0 & 0 & 0 & e^{-\frac{1}{3} (2 i \pi )} & 0 & 0 & 0 \\
 0 & 0 & 0 & 0 & 0 & e^{-\frac{1}{3} (2 i \pi )} & 0 & 0 \\
\end{array}
\right),\nonumber\\
&& I_x=\left(
\begin{array}{cccccccc}
 0 & 0 & 0 & 0 & 0 & 1 & 0 & 0 \\
 0 & 0 & 0 & 0 & 1 & 0 & 0 & 0 \\
 0 & 0 & 0 & 0 & 0 & 0 & e^{-\frac{1}{3} (2 i \pi )} & 0 \\
 0 & 0 & 0 & 0 & 0 & 0 & 0 & -1 \\
 0 & 1 & 0 & 0 & 0 & 0 & 0 & 0 \\
 1 & 0 & 0 & 0 & 0 & 0 & 0 & 0 \\
 0 & 0 & e^{\frac{2 i \pi }{3}} & 0 & 0 & 0 & 0 & 0 \\
 0 & 0 & 0 & -1 & 0 & 0 & 0 & 0 \\
\end{array}
\right),\quad R=\left(
\begin{array}{cccccccc}
 0 & 0 & 0 & 0 & 0 & 1 & 0 & 0 \\
 0 & 0 & 0 & 0 & 0 & 0 & e^{-\frac{1}{3} (2 i \pi )} & 0 \\
 0 & 0 & 0 & 0 & 1 & 0 & 0 & 0 \\
 0 & 0 & 0 & 0 & 0 & 0 & 0 & 1 \\
 0 & 1 & 0 & 0 & 0 & 0 & 0 & 0 \\
 0 & 0 & e^{\frac{2 i \pi }{3}} & 0 & 0 & 0 & 0 & 0 \\
 1 & 0 & 0 & 0 & 0 & 0 & 0 & 0 \\
 0 & 0 & 0 & 1 & 0 & 0 & 0 & 0 \\
\end{array}
\right).
\eea
This group has a small-group ID [ 288, 851 ]. The group is isomorphic to GL(2,3)$\times$S3.

We can use the same algorithm as in the previous section to classify the phase and group-subgroup relations. An example code is given in the attached file ``GAP$\_$kagome2.g''. We\ need to explain one more group theory concept.

The role of outer automorphisms: 
An automorphism is a map from the group to itself, respecting the group multiplication. 
Inner automorphism is given by conjugation $g_i \rightarrow h^{-1} g_i h$, with $h\in G$.
An outer automorphism is an automorphism that is not given by group conjugation. 
The outer automorphism group can be defined as the quotient of the full automorphism group by the inner automorphism group. For finite groups, such a coset is also a group. 
Suppose we have a phase that preserves the symmetry group $H$, then an outer automorphism may generate another phase preserving the symmetry group $H'$, and the two subgroups are not conjugate.
The two phases are therefore not equivalent.
To be more precise, the outer automorphism group can act on conjugate classes of subgroups. 
The conjugate classes of subgroups should form representations of the outer automorphism group. 
In our case, the outer automorphism group is given by C2$\times$C2.
To check this, we can calculate explicitly the full automorphism group and the inner automorphism group using the functions ``AutomorphismGroup()'' and ``InnerAutomorphismsAutomorphismGroup()'' respectively. The outer automorphism group can then be constructed using ``FactorGroup()''. 

Does outer automorphism change the subduction index? Possibly. One type of outer automorphism is given by the map 
\begin{align}
    g_i\rightarrow s^{-1}g_i s, \quad {\rm with} \quad s\notin G.
\end{align}
These outer automorphisms do not change the subduction index because $s^{-1}g_i s$ leave $s^{-1}.\vec{v}$ invariant, if $g.\vec{v}=\vec{v}$. There are also other outer automorphisms which do not belong to the above class. 
They will potentially change the subduction index.
For the Cubic group, we found that outer automorphisms indeed change the subduction index. 
For the group GL(2,3)$\times$S3 that we are working on in this subsection, we find that their outer automorphisms do not change the subduction index. In GAP, automorphisms are treated as a map, one can apply an automorphism to a group element by the function ``Image()''.

We found 21 phases, they group themselves into 12 irreps of the outer automorphism group. The outer automorphism group is isomorphic to C2$\times$C2.
In this section, we treat phases that belong to the same irrep of the outer automorphism group as a single object for the sake of simplicity in display. We emphasize, however, that they are really different phases.
\begin{table}[ht]
\begin{tabular}{|l|l|l|l|l|c|}
\hline
number of scalars & Group          & Generators                                                                                                & sub. ind. & deg. & outer auto. set size \\ \hline
0                 & Sym            & \{Tu, Mx, R\}                                                                                             & 0        & 1  & 1                    \\ \hline
1                 & D12            & \{ R\textasciicircum{}3, Mx, (R.Tu)\textasciicircum{}2 \}                                                 & 1        & 24   & 4                    \\ \hline
2                 & S3             & \{ R.Tu\textasciicircum{}-1.R\textasciicircum{}-1.Mx.Tu, R\textasciicircum{}3.Tu.R\textasciicircum{}-1 \} & 1        & 48   & 2                    \\ \hline
3                 & C6             & \{ Tu.R\textasciicircum{}3.Tu, (R.Tu)\textasciicircum{}2 \}                                               & 2        & 48   & 2                    \\ \hline
4                 & S3             & \{ R.Tu\textasciicircum{}-1.R\textasciicircum{}-1.Mx.Tu, (R.Tu)\textasciicircum{}2 \}                     & 2        & 48   & 2                    \\ \hline
5                 & S3             & \{ R.Tu\textasciicircum{}-1.Mx.Tu\textasciicircum{}2, (R.Tu)\textasciicircum{}2 \}                        & 2        & 48   & 2                    \\ \hline
6                 & C2 $\times$ C2 & \{ Tu.R\textasciicircum{}3.Tu, Mx \}                                                                      & 2        & 72   & 2                    \\ \hline
7                 & C3             & \{ R\textasciicircum{}3.Tu.R\textasciicircum{}-1 \}                                                       & 2        & 96   & 1                    \\ \hline
8                 & C3             & \{ (R.Tu)\textasciicircum{}2 \}                                                                           & 4        & 96   & 1                    \\ \hline
9                 & C2             & \{ Tu.R\textasciicircum{}3.Tu \}                                                                          & 4        & 144  & 2                    \\ \hline
10                & C2             & \{ Mx \}                                                                                                  & 4        & 144  & 1                    \\ \hline
11                & C2             & \{ R.Mx \}                                                                                                & 4        & 144  & 1                    \\ \hline
12                & $\{e\}$           & \{\}                                                                                                      & 8        & 288  & 1                    \\ \hline
\end{tabular}
\caption{Here ``outer auto. set size'' stands for the number of different phases connected by the outer auto morphism, we label these phases as a single phase for simplicity. }
\end{table}
\begin{figure}[ht]
\centering
\includegraphics[width=12cm]{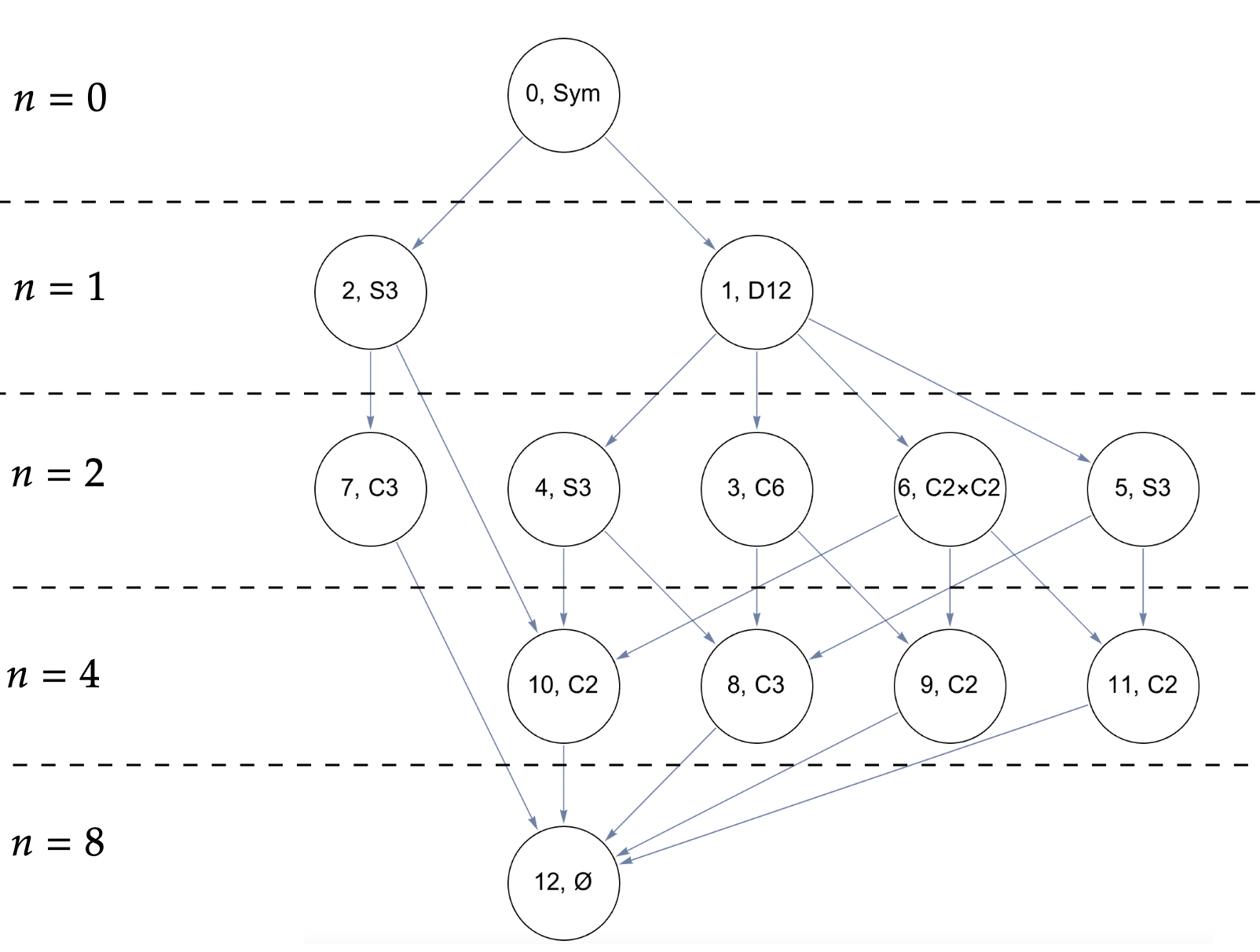}
\caption{Relations between different phases. The arrow indicates group-subgroup relations, which are possible 2nd-order phase transitions.}
\label{phases3}
\end{figure}
The relations between these phases are summarized in Fig.~\ref{phases3}.
Many of these phases were already discovered in~\cite{huh2011vison}.

\section{The quantum dimer/loop model (under construction)}

\begin{figure}
\centering
\includegraphics[scale=0.45]{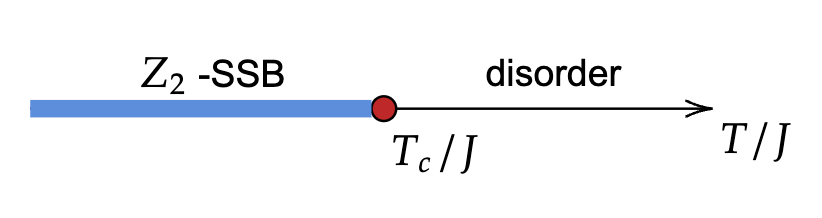}
\caption{The phase diagram of the transverse field Ising model.}
\label{QIsingphase}
\end{figure}

In addition to structural phase transitions, scalar CFTs can emerge from certain 2+1 dimensional lattice systems. In this case, the CFT corresponds to quantum critical points. We first consider the famous transverse field Ising model, with the Hamiltonian 
\begin{align}
    H=-J \sum_{\langle ij \rangle} Z_i Z_{j}+h \sum_i X_i
\end{align}
Here $X, Y$, and $Z$ are Pauli matrices. The index $i$ labels the lattice sites, let us just consider for example the square lattice. The operator 
$W=X_1\otimes \ldots \times X_N$
clearly commutes with the Hamiltonian, which is a global $Z_2$ symmetry.
When $h<<J$, we can neglect the second, the system has two group states 
$$
|+\rangle= |+\rangle_1 \otimes \ldots \otimes |+\rangle_N
$$
and 
$$
|-\rangle= |-\rangle_1 \otimes \ldots \otimes |-\rangle_N
$$
When $h>>J$, on the other hand, we can neglect the second, the ground state is 
$$
|\leftarrow \rangle=|\leftarrow \rangle_1 \otimes \ldots \otimes | \leftarrow \rangle_N
$$
Here $|\leftarrow \rangle_1=-\im |+ \rangle+ |-\rangle$ denotes the eigenvector of $X_1$. The term $h \sum_i X_i$ corresponds to a homogeneous magnetic field point along the x-direction. Notice also 
$$
\langle \leftarrow |Z_i |\leftarrow \rangle =0.
$$
The system is in the $Z_2$ symmetry unbroken phase, also called the disordered phase.
The model is the simplest model of quantum phase transition (phase transition that happens at zero temperature). It phase diagram is in Fig.~\ref{QIsingphase}.

How do we understand the phase transition in field theory language? 
We have discussed the Landau mean-field theory for spontaneous symmetry breaking in Section~\ref{Landautheory}. Including fluctuations, we get the action
\begin{align}
\mathcal{L}= \frac{1}{2}\bigg(\partial_{\mu}\phi(x, t) \bigg)^2 + m^2 \phi^2 +\frac{1}{4!} \lambda \phi^4.
\end{align}
Notice the first term involves derivatives along both the spatial and time directions. Since we are at zero temperature, the phase transition is driven by quantum fluctuations.
At finite temperature, when considering the physics at a scale much larger than the inverse temperature, we need to perform a Kaluza-Klein along the temporal circle~\cite{Chai:2020zgq},
\begin{align}
\phi(x,\tau)=\phi_0(x)+\sum_{n \in \mathbb{N}} \phi_{n}(x) e^{\im \frac{2\pi n}{\beta} \tau}+\phi^\dagger_{n}(x) e^{-\im \frac{2\pi n}{\beta} \tau}.
\end{align}
The action $S=\int d\tau \int d^3 x \mathcal{L}$ becomes $S= \int dx^3 \mathcal{L}',$ with (we set $m^2=0$)
\begin{align}
 \mathcal{L}'=\frac{1}{2}\partial_{i}\phi_0\partial_{i}\phi_0+\sum_{n \in \mathbb{N}} \partial_{i}\phi_n \partial_{i} \phi^\dagger_n+\sum_{n \in \mathbb{N}} \frac{4\pi^2 n^2}{\beta^2}\phi_n \phi^\dagger_n+ \frac{\lambda }{2\beta}\phi_0^2\sum_{n \in \mathbb{N}} \phi_n \phi^\dagger_n+\frac{\lambda}{4!\beta}\phi_0^4.
\end{align}
(We have redefined $\phi_n \rightarrow \frac{1}{\sqrt{\beta}} \phi_n$.) This is an interaction theory between one massless scalar and infinitely many massive scalar fields. 
The 1-loop diagram will renormalize the mass 
\begin{figure}
\centering
\includegraphics[scale=0.6]{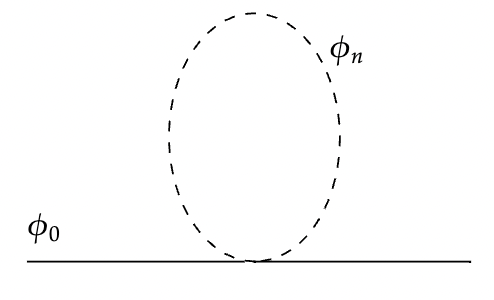}
\caption{The one-loop diagram which renormalizes the thermal mass.}
\end{figure}
The renormalized mass is 
\begin{align}
 m^2_{\rm th}=\sum_{n\in \mathbb{N}}\frac{\lambda}{\beta}\int\frac{d^3k}{(2\pi)^3} \frac{1}{k^2+(\frac{2\pi n}{\beta})^2}=-\frac{\lambda}{2\beta^2}\sum_{n\in \mathbb{N}} n = -\frac{\lambda}{2\beta^2} \zeta(-1)=\frac{\lambda}{24}\beta^{-2}.
\end{align}
We have used analytical continuation in dimension to regularize the integral, and Zeta function regularization to regularize the infinite sum.
The phase transition happens at $m_{\rm total}^2=m^2+ m^2_{\rm th}=0$. 
The mass $m^2$ corresponds to how far our system is away from the quantum critical point, that is $m^2= c_1 (h-h_c)+\cdots $, we therefore get 
$$c_1 (h-h_c)+ \frac{\lambda}{24} T_c^2=0,$$ 
which leads to
\begin{align}
T_c\sim |h-h_c|^{1/2}.
\end{align}
When we go beyond the mean-field theory approximation, the relations will be modified, and the thermal mass becomes
\begin{align}
m^2_{\rm thermal}=c~T^{1/\nu},
\end{align}
leading to 
\begin{align}
  T_c\sim |V-V_c|^{\nu}.
\end{align}
Here $\nu$ is a critical exponent.
In addition to the transverse Ising model, scalar CFTs appear in many quantum lattice systems. 
These include the Ising model \cite{Vidal:2008uy,tupitsyn2010topological,somoza2021self, Bonati:2024jqf}, the O(2) CFT \cite{sachdev1999translational,huh2011vison,YCWang2018}, the $N=3$ Cubic CFT \cite{PhysRevLett.61.2376,PhysRevB.92.075141,moessner2001resonating,Yan:2022zwt,Ran:2023wju,Ran:2024xcr}, and the O(4) CFT \cite{moessner2001ising,PhysRevLett.86.1881,yan2021topological}.
Among these lattice models, one interesting class is the quantum dimer/loop models, which are related to spin liquid, topological order, frustrated magnets, and Rydberg blockade.
The effective actions of these transitions are also closely related to the representations of space groups, more precisely, projective representations of the space group \cite{balents2005putting,Huh:2011pp}.  
The phase transitions are usually driven by an order parameter charged under a U(1) gauge symmetry (or its subgroup). 
The order parameter will experience a background gauge field on the lattice, which causes them  
to pick up a Berry phase when moving on the lattice. 
This explains the appearance of the projective representations~\cite{huh2011vison} of the space group. 

We will review various quantum dimer/loop models and their quantum field theory interpretations, focusing on the dynamics of the phase transition.

\subsection{the fully packed quantum loop model on the triangular lattice}\label{QLMandCubicCFT}
\begin{figure}[H]
\centering
\includegraphics[scale=0.5]{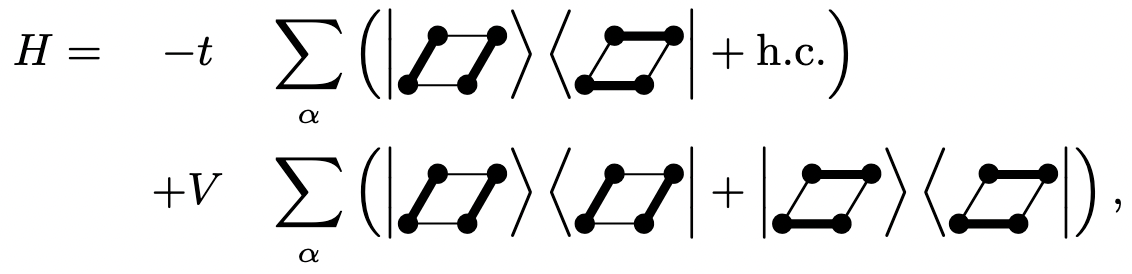}
\caption{The quantum dimer/loop model Hamiltonian}\label{dimerH}
\end{figure}

\begin{figure}[H]
\centering
a) \includegraphics[scale=0.6]{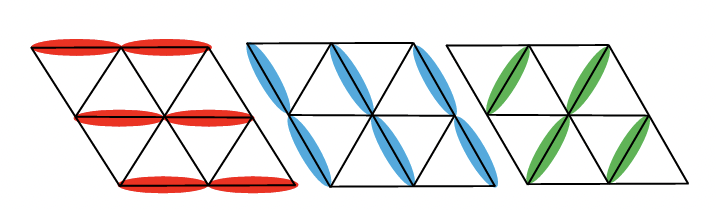} b)\includegraphics[scale=0.4]{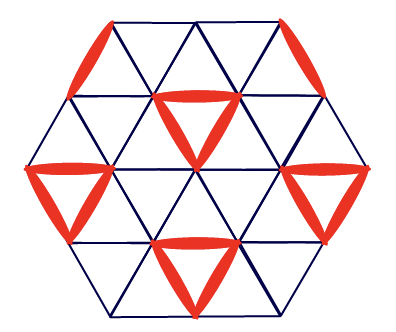}
\caption{a) The lattice nematic phase. b) The staggered phase. Image source:~\cite{PhysRevB.92.075141}.}\label{latticenematic}
\end{figure}

\begin{figure}[H]
\centering
\includegraphics[scale=0.4]{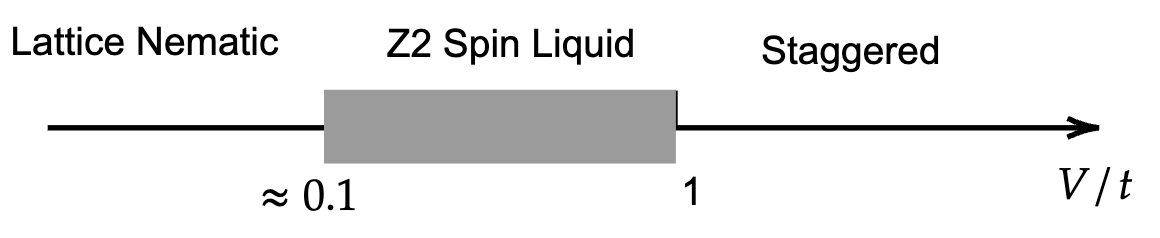}
\caption{Schematic phase diagram of the quantum loop model}\label{QLMphase1}
\end{figure}

We now discuss the fully packed quantum loop model, whose Hamiltonian is defined in Fig.~\ref{dimerH}.
The Hilbert space is given by all fully packed loop configurations on the triangular lattice. Another way to say it is that the Hilbert space is given by dimer configurations with two dimers per site local constraints. This is a generalization of the famous Rokhsar-Kivelson quantum dimer model with one dimer per site constraint in~\cite{PhysRevLett.61.2376,kivelson1987topology,moessner2001resonating,moessner2001ising}.
For the history of the dimer model, see~\cite{moessner2002resonating}.
The physics of the model is also very similar to the quantum dimer model, as discussed in~\cite{PhysRevB.92.075141}. When $V/t \ll 0$, the ground state is in the lattice nematic (LN) phase, see Fig.~\ref{latticenematic} a), as the first term in the Hamiltonian vanishes on these configurations. 
When $V/t \gg 0$, the ground state is in the staggered phase, see Fig.~\ref{latticenematic} b), as the second term in the Hamiltonian vanishes on these configurations.
There is another point where the Hamiltonian is exactly solvable, namely $V/t=1$, also called the Rokhsar-Kivelson (RK) point~\cite{PhysRevLett.61.2376}. 
The ground state is given by the equal amplitude superposition of all loop configurations.
At $V/t=1$, the Hamiltonian can be written as a positive semi-definite form. 
The state with equal amplitude superposition of all loop configurations clearly has zero energy and, therefore is the ground state, for more details, see~\cite{Ardonne:2003wa}.
We can draw a schematic phase diagram as in Fig.~\ref{QLMphase1}.
The phase transition from the Z2 spin-liquid phase to the lattice nematic phase is driven by the so-called vison degree of freedom. 
It is easy to map a dimer configuration on the triangular lattice to a vison configuration on the dual lattice.
We just need to follow the role that when passing a quantum loop, the vison changes sign. 
The lattice nematic phase can be viewed as a vison crystal.
\begin{figure}[H]
\centering
\includegraphics[scale=0.4]{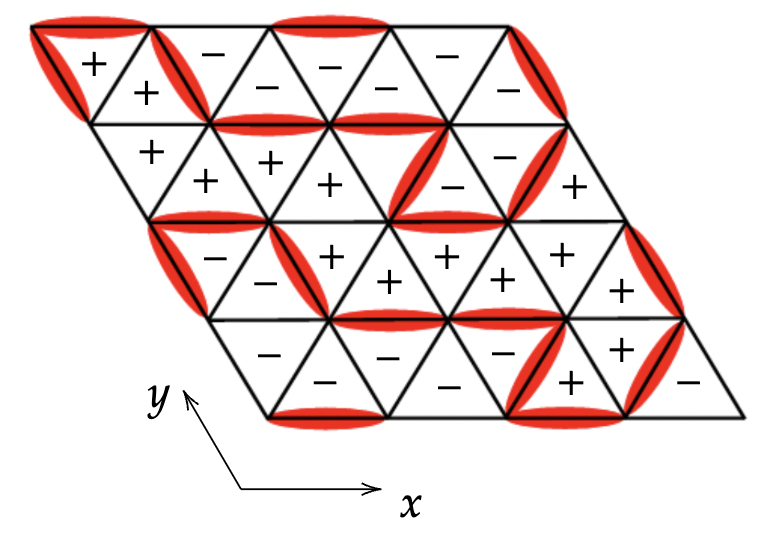}
\caption{a) The map from a dimer configuration to a vison a configuration.}\label{dimertoVison}
\end{figure}
The order parameter of the phase transition is given by
\begin{equation}\label{orderparameter}
	\begin{split}
		\phi_1=\sum_\mathbf{r}(v_{1,\mathbf{r}}+v_{2,\mathbf{r}})e^{i\pi x},\quad
		\phi_2=\sum_\mathbf{r}(v_{1,\mathbf{r}}+v_{2,\mathbf{r}})e^{i\pi y},\quad
		\phi_3=\sum_\mathbf{r}(v_{1,\mathbf{r}}-v_{2,\mathbf{r}})e^{i\pi (y-x)},
	\end{split}
\end{equation}
where $x$ and $y$ are the coordinates of all plaquettes of the triangular lattice, see Fig.~\ref{dimertoVison}.
This order parameter was first derived in~\cite{PhysRevB.92.075141}, following a similar calculation for the dimer model~\cite{moessner2001resonating,huh2011vison}.
These modes are located at the so-called ``M'' points of the Brillouin zone, which are points of symmetry as discussed in Section~\ref{Lifshitz}.
A more detailed discussion can be seen in supplement note 1 of \cite{Yan:2022zwt}. 
The quantum loop Hamiltonian induces interactions among the visons degree of freedom, which gives us a ferromagnetic transverse field Ising model on the dual lattice. 
In the weak coupling limit, one can analyze the dispersion relation, which helps us identify the critical mode \eqref{orderparameter} that drives the phase transition. 
\junchen{Discuss more details about the dual transverse field Ising model.}
The visons live on the dual honeycomb lattice, which can be divided into two sub-lattices.
The $(v_1(\vec{r}),v_2(\vec{r}))$ denotes visions living on the two sub-lattices respectively.
Similar to the structural phase transition discussed in Section~\ref{landautheoryandimage}, the order parameter forms an unfaithful (projective) representation of the space wallpaper group on the dual honeycomb lattice,
\begin{alignat}{5}
	T_x&=\left(
	\begin{array}{ccc}
		-1 & 0 & 0 \\
		0 & 1 &  0\\
		0 &  0 & -1\\
	\end{array}
	\right),\quad
	&&T_y&&=\left(
	\begin{array}{ccc}
		1 & 0 & 0 \\
		0 & -1 &  0\\
		0 &  0 & -1\\
	\end{array}
	\right),\quad
	&&\mathcal{I}&&=\left(
	\begin{array}{ccc}
		1 & 0 & 0 \\
		0 & 1 &  0\\
		0 &  0 & -1\\
	\end{array}
	\right),
	\label{I}\\
	R_6&=\left(
	\begin{array}{ccc}
		0 & 0 & -1 \\
		1 & 0 &  0\\
		0 &  1 & 0\\
	\end{array}
	\right),
	\label{r6}
\quad
	&&Z_2&&=\left(
	\begin{array}{ccc}
		-1 & 0 & 0 \\
		0 &-1 &  0\\
		0 &  0 & -1\\
	\end{array}
	\right),
\quad
	&&\mathcal{M}&&=\left(
	\begin{array}{ccc}
		0 & 0 & -1 \\
		0 &1 &  0\\
		-1 &  0 & 0\\
	\end{array}
	\right).
\end{alignat}
The above space group elements are the translations $T_x$ and $T_y$, the bond inversion $\mathcal{I}$, the six-fold rotation $R_6$, the mirror symmetry along the $y$ axis $\mathcal{M}$. 
The group generated by the above matrices is isomorphic to the cubic group.
The global $Z_2$ action reflects our freedom in choosing the overall sign of the visons, which is a gauge choice.  The corresponding Landau action is our favorite Cubic model,
\begin{align}\label{cubicCFT}
 \mathcal{L}=\frac{1}{2}\partial_{\mu} \phi^i \partial_{\mu} \phi^i +m^2 \phi^i\phi^i+u (\phi^i\phi^i)^2+v \sum_i (\phi^i)^4. 
\end{align}
At the RK point, the ground state is given by an equal amplitude superposition of all loop configurations, so that the visons are disordered. This corresponds to the disordered phase of \eqref{cubicCFT}, see Fig.~\ref{cubephases}.
What is the lattice nematic phase then? Mapping the loop configuration and then using \eqref{orderparameter}, we conclude that the LN phase corresponds to the face-center cubic phase of \eqref{cubicCFT}.
Minimizing the Landau potential (truncated to $\phi^4$ order), we notice that the theory is in the corner cubic phase when $v>0$, and in the face-centered cubic phase when $v<0$.

\begin{figure}[h]
\centering
\includegraphics[scale=0.4]{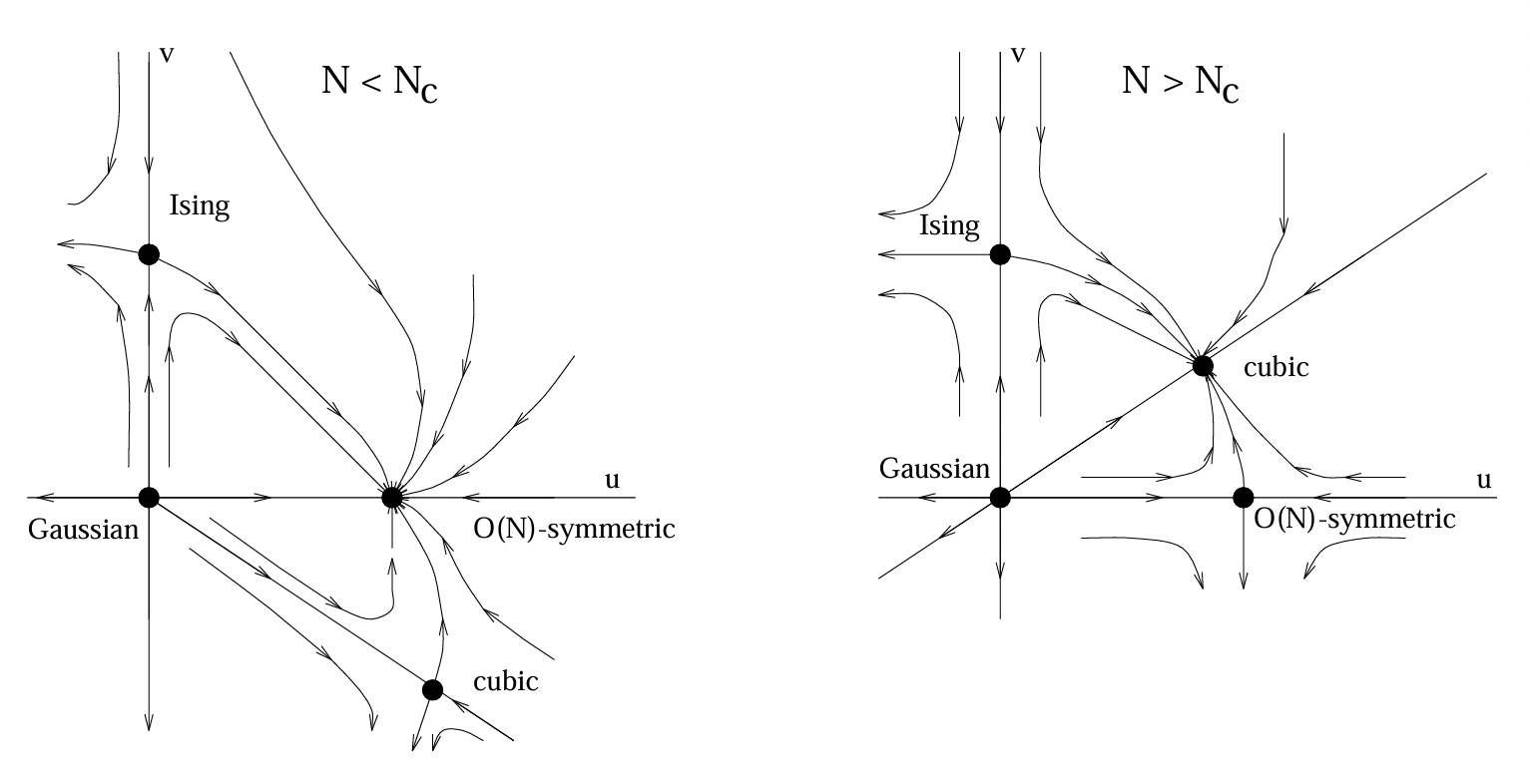}
\caption{RG flow of the Cubic theory below and above N$_c$. This result is obtained in perturbative RG calculation in $D=4-\epsilon$ expansion \cite{Aharony:2022ajv}. The conformal bootstrap study in~\cite{Chester:2020iyt} tells us that $N_c$ is slightly below 3.
Image source:~\cite{Pelissetto:2000ek}. }\label{CubicRG}
\end{figure}

The Cubic model was first introduced in \cite{aharony1973critical}. 
The model can be generalized to an arbitrary number of scalar fields. 
The two competing fixed points are the O($N$) invariant fixed point and the fixed point where with $S_N\ltimes (Z_2)^N$ symmetry, which we will call the $N$-state Cubic fixed point. 
The stability of the Cubic fixed point remained unclear for a long time (see Section 11.3 of \cite{Pelissetto:2000ek}).
As noted in \cite{aharony1973critical}, there exists a critical $N_c$, above which $N$-state Cubic fixed point becomes more stable than O($N$) invariant fixed point, see Fig.~\ref{CubicRG}.
To determine that $N_c$ has been the subject of many theoretical works, see \cite{Aharony:2022ajv} for a review of the early theoretical works.
The bootstrap result of~\cite{Chester:2020iyt} proves non-perturbatively that $N_c < 3$, so both the 3-state and the 4-state Cubic models should be non-perturbatively stable (see \cite{Rong:2017cow,Stergiou:2018gjj,Kousvos:2018rhl,Kousvos:2019hgc} for earlier works which attempts to bootstrap the Cubic CFT directly). 
Also, the 3-sate Cubic model can be studied using conformal perturbation theory~\cite{Rong:2023owx}, since the RG flow from the O(N) CFT to the Cubic is extremely short. 
See also~\cite{Hasenbusch:2022zur, Hasenbusch:2023fmn} for recent work on studying the Cubic CFT using Monte Carlo simulation of an improved $\phi^4$ model on the lattice. 

What does the above result imply for the quantum loop model? Notice the $N=3$ Cubic CFT is located in the $v>0$ region. This means that if the phase transition is 2nd order (through the Cubic CFT), the transition is from the disordered phase to the corner cubic phase. So the phase diagram in Fig.~\ref{QLMphase1} may need modification.

\begin{figure}[H]
\centering
\includegraphics[scale=0.4]{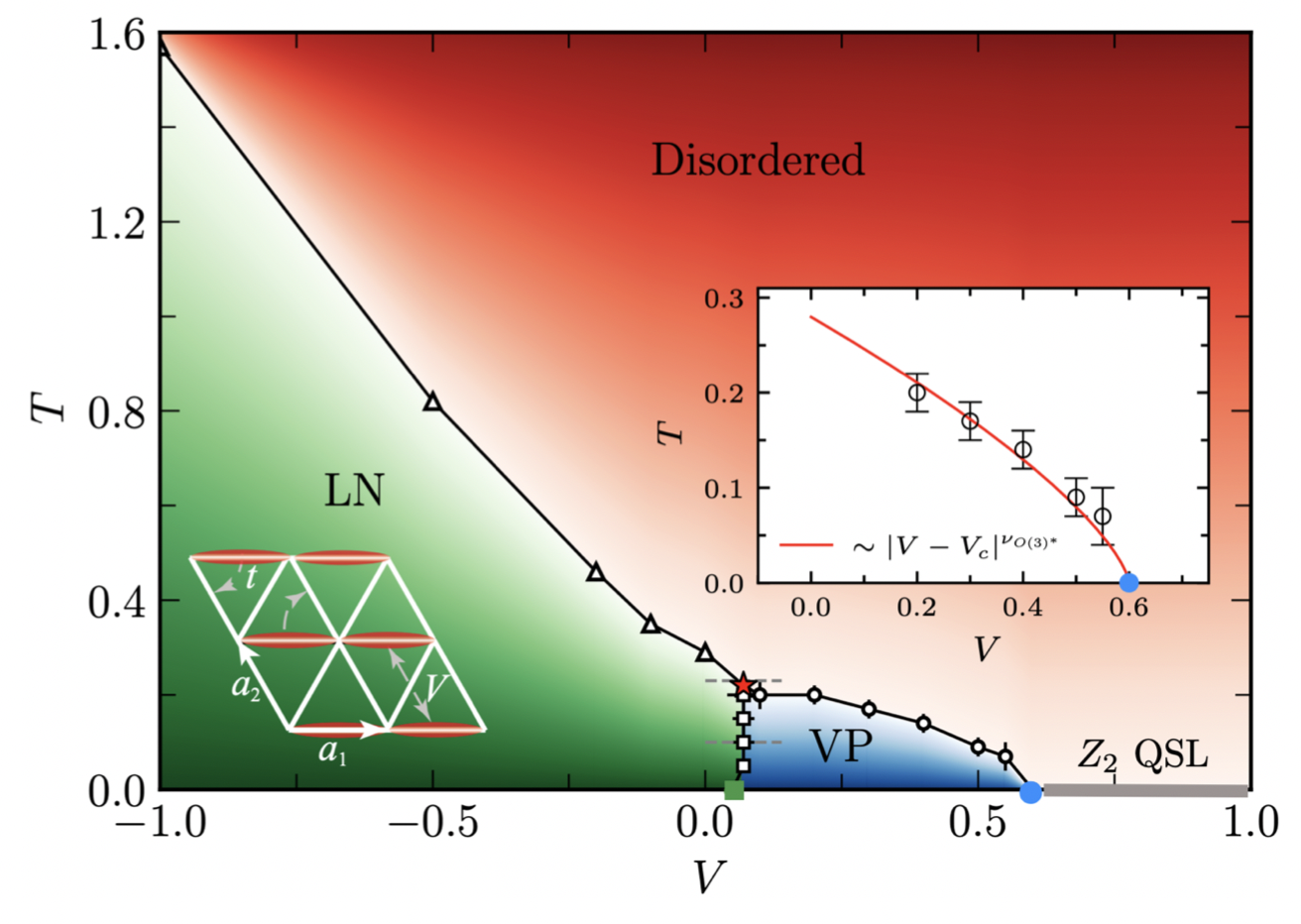}
\caption{The phase diagram of the quantum loop model from Quantum Monte Carlo simulation. The lattice nematic phase (LN) is the face-centered cubic phase of the Cubic model. The vison plaquette (VP) phase is the cornon cubic phase. Image source:~\cite{Ran:2024xcr}. Here $T$ denotes the temperature. (We have set $t=1$.)}\label{QLMphase2}
\end{figure}

Through a sequence of work using quantum Monte Carlo simulation~\cite{Yan:2022zwt,Ran:2023wju,Ran:2024xcr}, generalizing the earlier work in~\cite{PhysRevB.92.075141, PhysRevB.92.174402}, we obtained the phase diagram of the quantum loop model, as reproduced in Fig.~\ref{QLMphase2}. The newly discovered vison plaquette (VP) phase corresponds to the corner cubic phase.

\vskip 0.4 in
{\noindent\large  \bf Acknowledgments}
\vskip 0.in
This lecture note is based on a series of talks given by the author in a few places, including a journal club talk given at the Institut des Hautes  Études Scientifiques, and also lectures given in the Department of Physics, the University of Hong Kong. 
The author would like to thank the audience of these lectures and many other friends for valuable questions and comments, in particular, Slava Rychkov, Balt van Rees, Gregory Korchemsky, Jiaxin Qiao, Benoit Sirois, Marten Reehorst, Fidel Ivan Schaposnik, Zechuan Zheng and Aditya Hebbar, Martin Hasenbusch, Johan Henriksson, Stefanos Kousvos, Andreas Stergiou, Ziyang Meng, Weilun Jiang, Yang Qi, Xiaoyan Xu, and Fabien Alet. 
The manuscript is partially finished during the workshop ``Bootstrapping Nature: Non-perturbative Approaches to Critical Phenomena" and also the ``Hong Kong Computational and Theoretical Physics
Study Group 2023".
We thank the Galileo Galilei Institute and the University of Hong Kong receptively for their hospitality.
The 2nd version of this lecture note was updated while preparing a talk given for the \href{https://sites.google.com/view/thermalseminars}{``Thermal Seminars''} series, we thank the organizers for organizing such an event. 
The research of the author is supported by the Huawei Young Talents Program at IHES.
The author also acknowledges funding from the European Union (ERC “QFTinAdS”, project number 101087025).

\appendix 
\section{The Lifshitz condition in momentum space}\label{lifshitz}

{\it Quantum phase transitions.} Recently, it was noticed that certain 2+1 dimensional scalar CFTs can be realized in quantum phase transitions.
These include the Ising model \cite{Vidal:2008uy,tupitsyn2010topological,somoza2021self, Bonati:2024jqf}, the O(2) CFT \cite{sachdev1999translational,huh2011vison,YCWang2018}, the $N=3$ Cubic CFT \cite{PhysRevLett.61.2376,PhysRevB.92.075141,moessner2001resonating,Yan:2022zwt,Ran:2023wju,Ran:2024xcr}, and the O(4) CFT \cite{moessner2001ising,PhysRevLett.86.1881,yan2021topological}. 
These quantum transitions typically happen on two-dimensional lattices at zero temperature. 
The thermal fluctuations that drive the phase transition are replaced by quantum fluctuations. 
The effective actions of these transitions are also closely related to the representations of space groups, more precisely, projective representations of the space group \cite{balents2005putting,Huh:2011pp}.  
The phase transitions are usually driven by an order parameter charged under a U(1) gauge symmetry (or its subgroup). 
The order parameter will experience a background gauge field on the lattice, which causes them  
to pick up a Berry phase when moving on the lattice. 
This explains the appearance of the projective representations~\cite{huh2011vison} of the space group. 
It will be interesting to study these phase transitions systematically using group theory methods.

The Lifshitz condition is equivalent to the condition that the mass of the critical mode should be located at a local minimum in momentum space, that is 
\be{}
\frac{\partial a(T,P,\Vec{k})}{\partial k_1}=0,\quad 
\frac{\partial a(T,P,\Vec{k})}{\partial k_2}=0, \quad
\frac{\partial a(T,P,\Vec{k})}{\partial k_3}=0.
\ee
We review here a derivation given in \cite{toledano1987landau}. The free energy is a functional of the density function 
\be{}\label{freeE}
F[\rho_0+\delta\rho(\Vec{r})]=F[\rho_0]+\int d\Vec{r} d\Vec{r'}G(\Vec{r},\Vec{r'}) \delta\rho(\Vec{r})\delta\rho(\Vec{r'}).
\ee
The density fluctuation can be expand in irreps of the space group $G$ (for small $\Vec{q}$),
\be{}
\delta\rho(\Vec{r})=\sum_{\Vec{k}\in \Vec{k}_*}\sum_{\mathcal{\tilde{R}}} \sum_{i}^{dim({\mathcal{\tilde{R}}})} \int d\Vec{q} \phi^{\mathcal{\tilde{R}}}_{i}(\Vec{k}+\Vec{q}) \eta^{\mathcal{\tilde{R}}}_{i,\Vec{k}+\Vec{q}}(\Vec{r}).
\ee
In \eqref{isingdensity} and \eqref{structuremode2} we focused on a single mode in the expansion.
The vector $\Vec{k}$ is a point in the Brillouin zone and $\Vec{q}$ is a small deviation of the momentum. 
Here ${\mathcal{\tilde{R}}}$ is the irrep of the little group that keeps the vector $\Vec{k}$ invariant. 
The dimension of the irrep of the space group equals ${dim({\mathcal{\tilde{R}}})}$ times the number of vectors in $\Vec{k}_*$. 
We will drop the dependence on $k_*$ and $\mathcal{\tilde{R}}$ for simplicity.
The second term in \eqref{freeE} is therefore 
\be{}
F_2=\sum_{ij} \int d\Vec{q} \phi_{i}(\Vec{k}+\Vec{q})\phi_{j}(-\Vec{k}-\Vec{q}) A_{i,j}(\Vec{k}+\Vec{q}),
\ee
with the kernel 
\be{}
A_{i,j}(\Vec{k}+\Vec{q})=\int d\Vec{r}d\Vec{r'}G(\Vec{r},\Vec{r'})\eta_{i,\Vec{k}+\Vec{q}}(\Vec{r})\eta_{j,-\Vec{k}-\Vec{q}}(\Vec{r'}).
\ee
The kernel can be expanded in $\Vec{q}$,
\bea{}
A_{i,j}(\Vec{k}+\Vec{q})&=&A_{i,j}(\Vec{k})+\Vec{q}\cdot \Vec{B}_{i,j}(\Vec{k})+\ldots\nonumber\\
A_{i,j}(\Vec{k})&=&\int d\Vec{r}d\Vec{r'}\left(\eta_{i,\Vec{k}}(\Vec{r})\eta_{j,-\Vec{k}}(\Vec{r'})+\eta_{i,\Vec{k}}(\Vec{r'})\eta_{j,-\Vec{k}}(\Vec{r})\right)G(\Vec{r},\Vec{r'}),\nonumber\\
\Vec{B}_{i,j}(\Vec{k})&=&-\im \int d\Vec{r}d\Vec{r'} \vec{r} \left(\eta_{i,\Vec{k}}(\Vec{r})\eta_{j,-\Vec{k}}(\Vec{r'})-\eta_{i,\Vec{k}}(\Vec{r'})\eta_{j,-\Vec{k}}(\Vec{r})\right)G(\Vec{r},\Vec{r'}).
\eea
We used the relation $\eta_{i,\Vec{k}+\vec{q}}(\Vec{r})=e^{\im \vec{q}\cdot\vec{r} } \eta_{i,\Vec{k}}(\Vec{r})$.
Diagonalizing $A_{i,j}(\Vec{k})$ and pick the lowest eigen-value gives us the mass of the critical modes $a(T,P,\vec{k})$. 
To get a vanishing derivative, we need 
\be{}
\Vec{B}_{i,j}(\Vec{k})=0.
\ee
This is possible if $G(\Vec{r},\Vec{r'})$ takes certain special forms. 
However, if the term vanishes due to symmetry reasons, second-order phase transitions have a better chance of happening. 
The function $\left(\eta_{i,\Vec{k}}(\Vec{r})\eta_{j,-\Vec{k}}(\Vec{r'})-\eta_{i,\Vec{k}}(\Vec{r'})\eta_{j,-\Vec{k}}(\Vec{r})\right)$ lives in the $[{\cal R}\otimes {\cal R}]_a$ representation. 
Notice also
\be{}
\Vec{B}_{i,j}(\Vec{k})=-\im \int d\Vec{r}d\Vec{r'} (\vec{r}+\vec{t}) \left(\eta_{i,\Vec{k}}(\Vec{r})\eta_{j,-\Vec{k}}(\Vec{r'})-\eta_{i,\Vec{k}}(\Vec{r'})\eta_{j,-\Vec{k}}(\Vec{r})\right)G(\Vec{r},\Vec{r'}),
\ee
for arbitrary $\vec{t}$. 
Requiring that $\Vec{B}_{i,j}(\Vec{k})$ vanish for generic $G(\Vec{r},\Vec{r'})$ therefore gives us precisely the Lifshitz condition,
\be{}
\text{Lifshitz condition:} \qquad \mathcal{V} \notin [{\cal R}\otimes {\cal R}]_a. 
\ee

\section{Incommensurate phase transitions and the weak Lifshitz condition}\label{weakLifshitzsection}
\begin{figure}[ht]
\includegraphics[width=8cm]{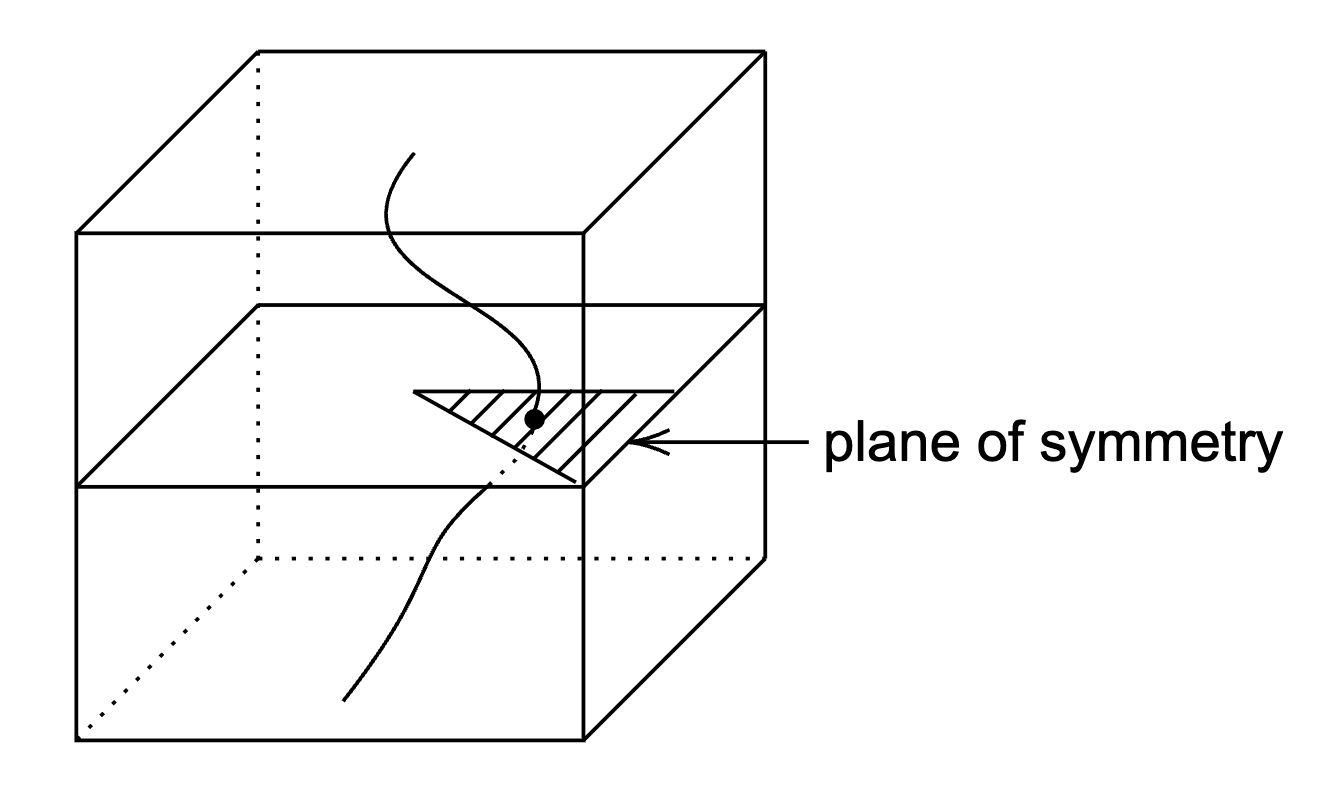}
\caption{Weak Lifshitz condition for planes of symmetry. }
\label{weakLifshitz}
\end{figure}

The incommensurate structural phase is controlled by a weaker version of the Lifshitz condition, which was introduced by Michelson in \cite{michelson1978weak}. 
The Brillouin zone contains points of symmetry, lines of symmetry, planes of symmetry, and generic points. 
For incommensurate transitions, the momenta of the critical mode are not located at points of symmetry. In general, the momentum is irrational numbers times the reciprocal lattice constant vector. 
This means the order parameter forms spatially modulated waves that are incommensurate with the lattice structure.
Let us denote the dimension of the symmetry domain that $\Vec{k}$ lives in as $m(\Vec{k})$. 
It can be proven that the number of Lifshitz invariants at $\Vec{k}$ is always bigger or equal to $m(\Vec{k})$, with $m(\Vec{k})=0,1,2,3$ for points of symmetry, lines of symmetry, planes of symmetry, and generic points respectively.
Take a plane of symmetry as an example, suppose the space group allows two Lifshitz invariants, the effective will contain two terms of the form \eqref{lifshitzterm}.
The two coupling constants of these terms,
\be
c_1(T,P,\Vec{k}), \quad c_2(T,P,\Vec{k}),
\ee
need to be zero for the transition to be second order.
The conditions $c_1(T,P,\Vec{k})= c_2(T,P,\Vec{k})=0$ correspond to a line in the Brillouin zone, which might intersect with the plane of symmetry at isolated points: let us denote the intersection point as $\Vec{k}_0$.
The location of $\Vec{k}_0$ depends on temperature and pressure $(T,P)$. 
We demonstrate the weak Lifshit condition for planes of symmetry in Figure \ref{weakLifshitz}.
The mass gap of this $\Vec{k}_0$ model is therefore a function of $(T,P)$, 
\be
a(T,P)=a\left(T,P,\Vec{k}_0(T,P)\right).
\ee
The second order phase transition line corresponds to $a(T,P)=0$, which is a one-dimensional line in the $(T,P)$ plane, therefore can be reached without fine-tuning. 
Unlike the points of symmetry which allow no Lifshitz invariants, the lines of symmetries allow at most one Lifshitz invariant, the planes of symmetry allow two Lifshitz invariants, and a generic momentum point allows three Lifshitz invariants. 
In summary, the number of allowed Lifshitz invariants should be equal to $m(\Vec{k})$. 
The irreps of the 230 crystallographic space that satisfy these weak Lifshitz conditions are classified in \cite{stokes1993landau}.

\bibliographystyle{JHEP}
\bibliography{reference.bib}

\end{document}